\documentclass[sigconf]{acmart}

\AtBeginDocument{%
  }

\setcopyright{acmcopyright}\acmConference[ESEM '22]{ACM / IEEE International Symposium on Empirical Software Engineering and Measurement (ESEM)}{September 19--23, 2022}{Helsinki, Finland}
\acmBooktitle{ACM / IEEE International Symposium on Empirical Software Engineering and Measurement (ESEM) (ESEM '22), September 19--23, 2022, Helsinki, Finland}
\copyrightyear{2022}
\acmYear{2022}
\acmDOI{10.1145/3544902.3546244}

\acmPrice{15.00}
\acmISBN{978-1-4503-9427-7/22/09}

\begin{document}

\title{Investigating the Impact of Continuous Integration Practices on the Productivity and Quality of Open-Source Projects}

\author{Jadson Santos}
\affiliation{%
  \institution{Federal University of Rio Grande do Norte}
  \city{Natal}
  \country{Brazil}
}
\email{jadson.santos@ufrn.br}

\author{Daniel Alencar da Costa}
\affiliation{%
  \institution{University of Otago}
  \city{Dunedin}
  \country{New Zealand}
}
\email{danielcalencar@otago.ac.nz}

\author{Uirá Kulesza}
\affiliation{%
 \institution{Federal University of Rio Grande do Norte}
 \city{Natal}
 \country{Brazil}
}
\email{uira@dimap.ufrn.br}

\renewcommand{\shortauthors}{Santos et al.}

\begin{abstract}
  \textbf{\textit{Background:}} Much research has been conducted to investigate the impact of Continuous Integration (CI) on the productivity and quality of open-source projects. Most of studies have analyzed the impact of adopting a CI server service (e.g, Travis-CI) but did not analyze CI sub-practices.
  \textbf{\textit{Aims:}} We aim to evaluate the impact of five CI sub-practices with respect to the productivity and quality of GitHub open-source projects. \textbf{\textit{Method:}} We collect CI sub-practices of 90 relevant open-source projects for a period of 2 years. We use regression models to analyze whether projects upholding the CI sub-practices are more productive and/or generate fewer bugs. We also perform a qualitative document analysis to understand whether CI best practices are related to a higher quality of projects. \textbf{\textit{Results:}} Our findings reveal a correlation between the Build Activity and Commit Activity sub-practices and the number of merged pull requests. We also observe a correlation between the Build Activity, Build Health and Time to Fix Broken Builds sub-practices and number of bug-related issues. The qualitative analysis reveals that projects with the best values for CI sub-practices face fewer CI-related problems compared to projects that exhibit the worst values for CI sub-practices. \textbf{\textit{Conclusions:}} We recommend that projects should strive to uphold the several CI sub-practices as they can impact in the productivity and quality of projects.
\end{abstract}



\maketitle

\section{Introduction}
Continuous Integration (CI) is a software development practice that aims to reduce the complexity of integrating code produced by different developers. It can also decrease the risks \cite{Duvall2007} of software projects by promoting the automated build and testing of their code. The automated process of CI can bring benefits such as the improvement in the delivery time of a new software release as well as decreasing the number of errors potentially generated by manual tasks. The adoption of CI requires the adoption of a set of sub-practices \cite{FowlerCI2006, Duvall2007, STAHL201448}, such as to perform code commits frequently, build the software frequently and develop and perform automated tests.

Over the last years, researchers have empirically studied CI from different perspectives \cite{EliezioRevisaoSistematica2022} \cite{StakeholderPerceptions2015} \cite{ReplicationCIPainPoints2019} \cite{TradeOffsCI2017} \cite{GustavoSizilio2019}. Recent research works have investigated the impact of CI on productivity and quality of software projects \cite{VasilescuQualityProdOutcomes2015} \cite{JoaoHelis2018}. Vasilescu et al. \cite{VasilescuQualityProdOutcomes2015} analyzed a dataset of 246 \textsc{GitHub} open-source projects that adopted \textsc{Travis-CI} at some point in their lifetime. They found that after adopting \textsc{Travis-CI}, core developers in those projects have significantly increased the number of merged pull requests (PRs) as well as discovered more bugs-related issue reports (IRs). Helis et al. \cite{JoaoHelis2018} conducted an empirical study that analyzed 162,653 PRs of 87 \textsc{GitHub} projects to evaluate the impact of adopting CI on the delivery time of merged PRs. They found a large increase in the number of submitted, merged and delivered PRs per release after CI was adopted. These empirical studies derive their results based on projects that have adopted a CI service, such as \textsc{Travis-CI}.\footnote{https://travis-ci.org/} Nevertheless, Felidré et al.~\cite{CITheater2019} observed that many projects that adopt a CI service, do not necessarily follow fundamental CI practices. They refer to this problem as \textit{Continuous Integration Theater}~\cite{CITheater2019}. CI Theater occurs when a project has adopted a CI service to automate their build and testing process but, on the other hand, do not dedicate much attention to other existing CI sub-practices, such as frequent code commits, frequent builds and automated tests.

In this context, this paper investigates the impact of different CI sub-practices on the productivity and quality outcomes of open-source projects. Similar to previous works \cite{VasilescuQualityProdOutcomes2015} \cite{JoaoHelis2018}, we define productivity in terms of merged PRs and quality in terms of closed bug-related IRs. Beyond selecting projects solely on the basis that they have used a CI service, we extract 5 CI sub-practices (build duration, build activity, build health, time to fix a broken build, and commit activity) and investigate the correlation of those sub-practices with the productivity (merged pull requests) and quality (closed bug-related issue reports) of a software project. Furthermore, we perform a qualitative analysis to investigate how much these CI sub-practices can reflect the maturity of the projects concerning the adoption of CI. We use a document analysis approach to analyze and compare the comments of PRs from two groups of projects - the \textit{Best CI Sub-Practices Group} and the \textit{Worst CI Sub-Practices Group}. Thereby, we asked the following research questions:

\begin{itemize}
\item \textbf{RQ1: Which CI sub-practices contribute to higher productivity outcomes?} We apply linear regression models to analyze the influence of CI sub-practices in the productivity of projects. Can a specific CI sub-practice increase the productivity of a project? 
\item \textbf{RQ2: Which CI sub-practice contribute to higher quality outcomes?} We apply linear regression models to analyze the influence of CI sub-practices in the quality of projects. Can a specific CI sub-practice increase the quality of a project?
\item \textbf{RQ3: Can consistently applied CI sub-practices be an indicator of process quality?} The goal of this research question is to verify whether the projects that best follow the 5 CI sub-practices are more mature regarding CI. Do developers face fewer problems in the development process than the developers of projects that have the worst values for CI sub-practices?
\end{itemize}

Based on our results, we find a positive correlation between 2 CI sub-practices and a increase in the number of merged PRs. We find also a positive correlation between frequency of builds and the increase of bugs-related IRs and our model shows that projects with poor build health tend to generate more bug-related IRs. We find also evidence that projects that uphold the CI sub-practices can indicate a better quality in the development process, while they face fewer problems regarding CI, compared to projects that do not uphold the CI sub-practices.

\textbf{Paper organization}. The rest of this paper is organized as follows: In Section 2, we explain the design of our empirical study, describing the motivation and methodology used in each RQ. In Section 3 we present the results of our empirical study, while we discuss the impact of these results in Section 4. In Section 5 we discuss the threats to the validity. In Section 6, we discuss the related work. Finally, we concluded the paper in Section 7.

\section{Study Settings \& Research Questions}
In this section, we first describe how we selected our projects and then we explain the motivation and methodology of each research question (RQ).

\subsection{Projects Selection}

To select the projects of our study, we searched for popular \textsc{GitHub} projects that use \textsc{Travis-CI} as their CI service. The projects must also have had a substantial number of builds, pull requests (PRs), issue reports (IRs) and commits. Lastly, our studied projects would need to be active (in terms of builds, PRs, IRs and commits) to perform our investigations. Similar to other studies \cite{JoaoHelis2018,VasilescuQualityProdOutcomes2015,Vassallo01}, we selected projects using \textsc{Travis-CI} as \textsc{Travis-CI} is one of the most popular CI services on GitHub \cite{VasilescuQualityProdOutcomes2015}. Differently from other CI servers (e.g., Jenkins), \textsc{Travis-CI} provides the entire build history of a project. Lastly, although \textsc{Github-Actions} has become popular, its addition to \textsc{GitHub} is still fairly recent,  \footnote{https://github.blog/2019-08-08-github-actions-now-supports-ci-cd/} which could hinder our goal of finding a 2-year history of builds in our projects. 

\begin{figure}[h]
  \centering
  \includegraphics[width=\linewidth]{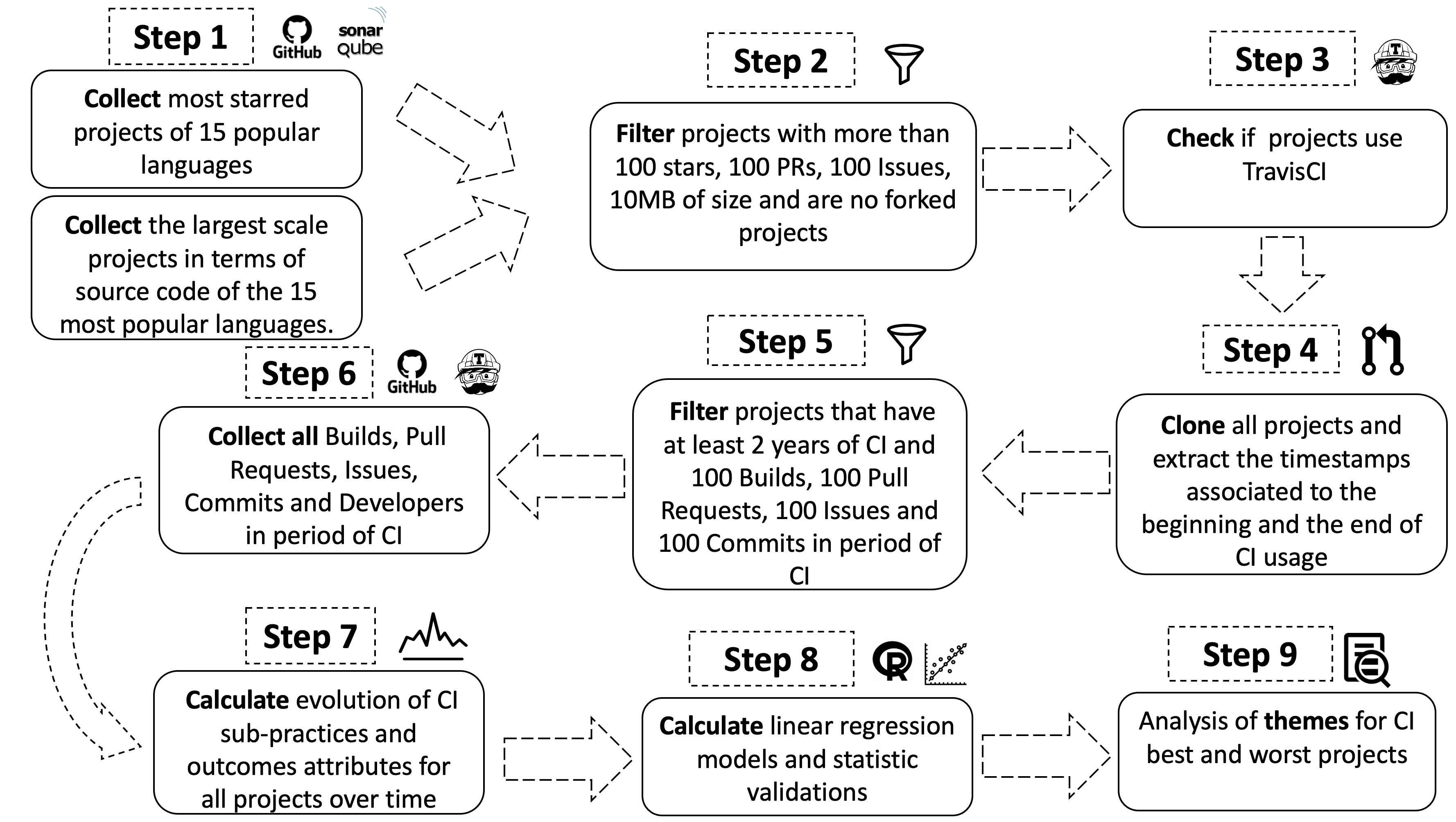}
  \caption{An overview of our methodology}
  \label{fig:overall_approach}
\end{figure}

Figure~\ref{fig:overall_approach} provides an overview of all steps involved in our project selection approach.

Our selection strategy was inspired by Zhao et al.'s work~\cite{VasilescuImpactCI2017}. In \textbf{Step 1}, we collected the most starred repositories for a set of 15 popular programming languages on GitHub: Java, JavaScript, C\#, Python, PHP, TypeScript, C, Go, C++, Kotlin, Ruby, Rust, Swift, Scala and Objective-C using the \textsc{GitHub} Search API,\footnote{https://docs.github.com/en/rest/reference/search\#search-repositories} yielding 11,671 projects. Next, to increase the number of projects in our study, we searched for additional projects on \textsc{SonarQube} using its web API.\footnote{https://community.sonarsource.com/t/list-of-all-public-projects-on-sonarcloud-using-api/33551} We only kept \textsc{SonarQube} projects that were also available on \textsc{GitHub}, obtaining an additional set of 16,212 projects. In total, we collected 27,883 projects from the two searches. 

\begin{figure}[h]
  \centering
  \includegraphics[width=\linewidth]{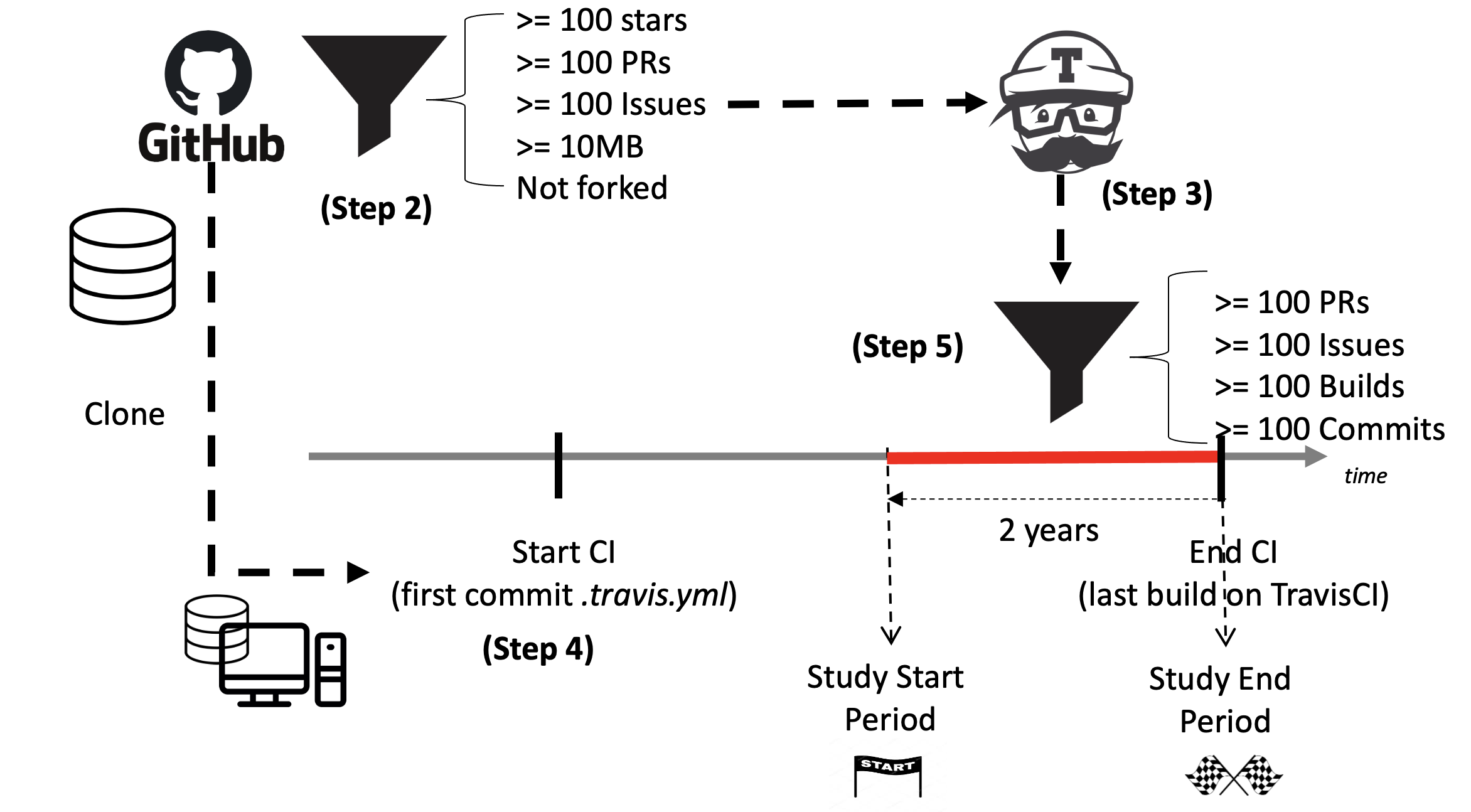}
  \caption{Overview of filtering non-relevant projects}
  \label{fig:project_selection}
\end{figure}

In \textbf{Step 2}, our goal was to eliminate non-relevant projects (e.g., toy projects). Considering the initial 27,883 projects, we first verified whether they survived the following thresholds: more than 100 stars, 100 PRs, 100 Issues, 10MB of size and were not forked projects. Figure~\ref{fig:project_selection} depicts the selection process for steps 2, 3, 4 and 5.  

The number ``100'' in our thresholds was inspired by previous studies \cite{VasilescuQualityProdOutcomes2015,JoaoHelis2018}. Additionally, similarly to prior research works \cite{MarcosOliverira2017,DEOLIVEIRA2019110420,ConvolutionalAttentionNetwork2017}, we used \textit{10MB of size} as a threshold to eliminate starred projects that do not represent meaningful projects. For example, the hello-world\footnote{https://github.com/octocat/hello-world} \textsc{GitHub} project has 1.7K stars, more than 700 Issues and 300 PRs, consisting only of a single README.md file for a demo.

After applying these first filters, 3,143 projects remained. From this set of projects, in \textbf{Step 3}, we checked whether they have used \textsc{Travis-CI}. We did so by using a URL pattern match \cite{Vassallo01} extracted from the project's name on \textsc{GitHub}. 2,029 projects remained. 

In \textbf{Step 4}, similarly to Zhao et al.'s  work~\cite{VasilescuImpactCI2017}, we cloned the \textsc{GitHub} repositories of all 2,029 projects. We identified the main \textsc{Travis-CI} branch of the repositories by identifying the earliest commit to the \texttt{.travis.yml} configuration file. We used the timestamp of the first commit to the \texttt{.travis.yml} file as the starting date of the CI usage within the projects. Afterwards, we retrieved the timestamp of the last build of the projects from their \textsc{Travis-CI} service. We considered the timestamp of the last build as the ending date of CI usage in the projects.  

We filter for projects that have at least \textbf{2 years} of activity on \textsc{Travis-CI} in \textbf{Step 5}. We obtain a total of 1,751 projects. Next, we performed a new filtering step, selecting projects with at least 100 Builds, 100 Pull Requests, 100 Issues and 100 Commits, but now considering the period of 2 years of \textsc{Travis-CI} activity. We obtained a total of 776 relevant projects. 

In \textbf{Step 6}, we collected information from PRs, IRs, commits, and builds from both \textsc{GitHub} and \textsc{Travis-CI} within the period of 2 years of CI usage. We collected only closed PRs and IRs. Regarding to closed PRs, we only analyzed merged PRs, i.e., PRs that have a non-null value within the \textit{merged\_at} field as retrieved by the \textsc{GitHub} API. In total, considering the 776 relevant projects, we collected 403,403 PRs, 983,460 IRs, 394,063 builds and 1,486,429 commits.

\begin{figure}[h]
  \centering
  \includegraphics[width=\linewidth]{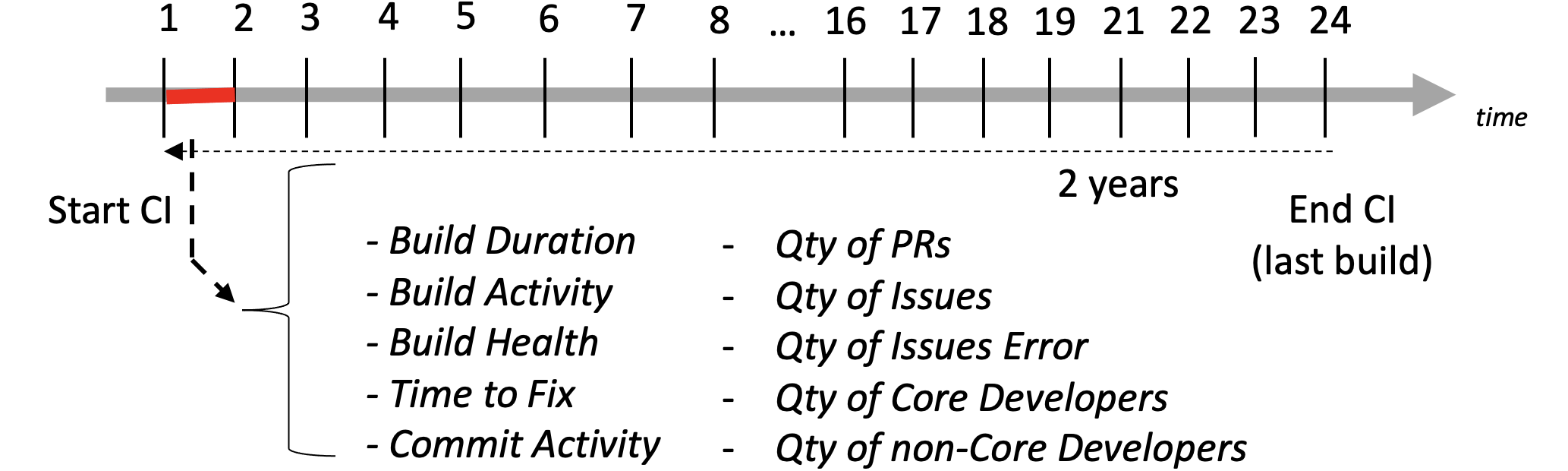}
  \caption{Periods of Analysis}
  \label{fig:periods_of_analysis}
\end{figure}

In the \textbf{Step 7}, we analyze the evolution of CI sub-practices (explained in Section~\ref{sec:cisubpractices}) within all the 776 relevant projects over time. To do so, we divided the project history into time intervals. Similar to Zhao et al.'s work~\cite{VasilescuImpactCI2017} and  Vasilescu et al.'s work~\cite{VasilescuQualityProdOutcomes2015}, we use an interval of \textbf{1 month} to segment our projects' history into periods of analysis. Table~\ref{tab:outcomes_attributes} shows the computed attributes with their respective definitions. Instead of using the period of 24 months centered around the adoption of \textsc{Travis-CI} (as in prior studies~\cite{VasilescuImpactCI2017}), to pursue our goals, we analyze the latest 24 months of CI usage in our projects. Our goal is to evaluate CI projects that have better recent values for the CI sub-practices. The latest 24 months is the period where projects should have had the most stable values for the CI sub-practices. Figure~\ref{fig:periods_of_analysis} illustrates the information collected for each period of analysis.

After segmenting our projects' history in periods, we noted that several projects had several periods without activity, i.e., 0 builds and 0 commits in a specific month. To only consider projects that remained active during the period of analysis, we filtered for projects that had at least 20 periods (80\% of the total number of periods) with at least 1 build and 1 commit. Thus, we have a final set of \textbf{90 relevant and active projects} used in our study.

\begin{table*}[htbp]
  \caption{Projects outcomes attributes collected for each of 24 months of study}
  \label{tab:outcomes_attributes}
  \begin{center}
  \begin{tabular}{lr} 
    \hline
\textbf{Atributes}            & \multicolumn{1}{l}{\textbf{Definition}}            \\ 
\hline

\textbf{qty\_pull\_request}& \multicolumn{1}{l}{number of PRs in the period}\\

\textbf{qty\_pull\_request\_core}& \multicolumn{1}{l}{Number of PRs in the period associated with core developers}\\

\textbf{qty\_pull\_request\_non\_core}& \multicolumn{1}{l}{Number of PRs in the period associated with non-core developers}\\

\textbf{qty\_developers\_prs} & \multicolumn{1}{l}{ Number of distinct PR authors in period }\\

\textbf{qty\_core\_developers\_prs} & \multicolumn{1}{l}{ Number of distinct PR core developers authors in period}\\

\textbf{qty\_non\_core\_developers\_prs} & \multicolumn{1}{l}{ Number of distinct PR non-core developers authors in period}\\

\textbf{qty\_issue}&\multicolumn{1}{l}{number of Issues in the period} \\

\textbf{qty\_issue\_core}&\multicolumn{1}{l}{Number of Issues in the period associated with core developers} \\

\textbf{qty\_issue\_non\_core}&\multicolumn{1}{l}{Number of Issues in the period associated with non-core developers} \\

\textbf{qty\_issue\_error}&\multicolumn{1}{l}{Number of Issues in the period associated with error labels} \\

\textbf{qty\_issue\_error\_core}&\multicolumn{1}{l}{ \begin{tabular}[c]{@{}l@{}} Number of Issues in the period associated \\ with error labels and core developers \end{tabular} } \\

\textbf{qty\_issue\_error\_non\_core}&\multicolumn{1}{l}{ \begin{tabular}[c]{@{}l@{}} Number of Issues in the period associated \\ with error labels and non-core developers \end{tabular} } \\

\textbf{qty\_developers\_issues}&\multicolumn{1}{l}{ number of distinct Issues authors in period} \\

\textbf{qty\_core\_developers\_issues}&\multicolumn{1}{l}{Number of distinct Issues core developers authors in period} \\

\textbf{qty\_non\_core\_developers\_issues}&\multicolumn{1}{l}{Number of distinct Issues non-core developers authors in period} \\

\hline
\end{tabular}
\end{center}
\end{table*}

In the last two steps of our method, we built our linear regression models \textbf{(Step 8)} to verify the potential influence of each CI sub-practice in the {\em productivity} and {\em quality}~\cite{VasilescuQualityProdOutcomes2015} outcomes of our projects. In \textbf{Step 9}, we perform a qualitative analysis to verify whether projects that uphold the CI sub-practices also show signs of quality in their development process. All collected information as well as the database and scripts used in this study are available in our online appendix.\footnote{\url{https://zenodo.org/record/6513155}}

\subsection{CI sub-practices}\label{sec:cisubpractices}

Existing research still lacks a well-defined set of criteria to analyze the potential impact of CI on software development~\cite{EliezioRevisaoSistematica2022}. For instance, several studies consider the use of a CI service (e.g., \textsc{Travis-CI}) as the sole criterion to indicate whether projects use CI or not~\cite{EliezioRevisaoSistematica2022}. However, Duvall et al.\cite{Duvall2007} and Fowler~\cite{FowlerCI2006} have defined a set of CI sub-practices that, combined together, can be used to determine whether a project uses CI or not. Additionally, Felidré et al. \cite{CITheater2019} revealed that many CI projects do not necessarily follow CI sub-practices consistently (e.g., infrequent commits are frequent). Therefore, in our work, instead of investigating whether the usage of \textsc{Travis-CI} can be associated with better quality and productivity outcomes~\cite{VasilescuImpactCI2017}, we study \textbf{5 CI sub-practices} in a more separate fashion to understand their potential contribution to quality and productivity outcomes.  

\begin{itemize}
\item \textbf{Build Duration}~\cite{CITheater2019,DanielBuildDuration2019} measures the duration of the build \textit{(build finished at timestamp - build started at timestamp)}. The build duration was retrieved directly from \textsc{Travis-CI}. To fit our models, we used the median build duration per period of analysis (1 month).

\item \textbf{Build Activity}~\cite{Duvall2007} is a unit interval (i.e., a closed interval [0,1]) representing the rate of builds across days, i.e, if builds were made every day in the period of analysis, the value would be 1. If builds were made in half of the days, the value would be 0.5. If there were no builds, the value would be 0. 

\item \textbf{Build Health}~\cite{EliezioRevisaoSistematica2022} is a unit interval representing the rate of build failures across days. If there were build failures every day, the value would be 0. if there were no build failures, the value would be 1.

\item \textbf{Time to Fix a Broken Build}~\cite{CITheater2019} consists of the median time in a period (1 month) that builds remained broken. When a build breaks, we compute the time in seconds until the build returns to the \textit{"passed"} status. If the analysis period ends and the build did not return to the \textit{"passed"} status, we consider the time since it was broken until the end of the period. When a period has no broken builds, the value would be 0.

\item \textbf{Commit Activity}~\cite{CITheater2019} is a unit interval representing the rate of commits across days. If commits were made every day in a period (1 month), the value would be 1. If commits were made in half of days the value would be 0.5. If there were no commits in a period the value would be 0.
\end{itemize}

In this paper, we focus on investigating the build and commit related sub-practices. These sub-practices were chosen because they cover most of the practices defined by Duvall et al \cite{Duvall2007} and Fowler \cite{FowlerCI2006} as well as they cover most of the sub-practices analyzed by Felidré et al \cite{CITheater2019}. The ``Code Coverage'' sub-practice analyzed by Felidré et al. \cite{CITheater2019} was initially considered in this study. However, we found few projects that collect and store coverage data for a long period of time on a centralized platform like \textsc{SonarQube}.\footnote{https://www.sonarqube.org/}

\subsection{Research Questions}

\subsubsection*{RQ1: Which CI sub-practices contribute to higher productivity outcomes?}

\hfill \break

\textbf{\textit{\underline{Motivation:}}} 
Vasilescu et al.~\cite{VasilescuQualityProdOutcomes2015} observed that CI is associated with higher productivity outcomes (i.e., number of merged PRs). However, we still do not know whether the increase in productivity is associated to only certain sub-practices of CI (as opposed to all of them). Better understanding which CI sub-practices have stronger associations with productivity outcomes (i.e., number of merged PRs~\cite{VasilescuQualityProdOutcomes2015}) may help us to optimize the usage of CI.

\textbf{\textit{\underline{Approach:}}} We fit linear regression models (Ordinary Least \\*Squares, OLS \cite{OLS2022}), to find associations between the number of merged PRs (Table~\ref{tab:outcomes_attributes}) and the 5 CI sub-practices. We would expect that the number of merged PRs should be higher in projects where CI sub-practices are employed consistently.

Because the number of developers in a project can impact the number of merged PRs, we control the influence of the number developers in our models. We create another model where we divide the number of merged PRs by the number of active developers in each period of analysis. 
Inspired by the findings of Vasilescu et al.'s work \cite{VasilescuQualityProdOutcomes2015}, we decided to divide our analysis in two groups of developers: (i) core developers and (ii) non-core developers. Our goal was to minimize the effects of the growth of internal and external contributors on our models.

To identify core developers, we combined strategies used by Vasilescu et al.~\cite{VasilescuQualityProdOutcomes2015} and Poncin et al.~\cite{MiningRepositories2011}. A core developer is a developer who: (i) \textit{``either had write access to a project’s code repository or had closed issues and pull requests submitted by others''} \cite{VasilescuQualityProdOutcomes2015} or (ii) \textit{``has been involved in the project for a relatively long period of time and has more revisions in the version control system than average''}~\cite{MiningRepositories2011}.

As it is common for open-source developers to use different aliases when performing contributions, we used a strategy to identify core developers from author information. Differently from Zhao et al.~\cite{VasilescuImpactCI2017} who stated: \textit{``we used heuristics that match first and last names, email prefixes, and email domains''}, after collecting such information from \textsc{GitHub}, we noticed that emails were not informed by many developers. Thus, we only used the first and last name of developers. To identify whether two developers have similar names, we applied a combination of 2 algorithms: (i) Jaro Winkler Similarity~\cite{JaroWinklerSimilarity2020} and (ii) Levenshtein Distance~\cite{LevenshteinDistance2017}. We empirically found that, with Jaro Index \textbf{$\geq$ 0.85} and Levenshtein Distance \textbf{$<$ 5}, we could accurately compare developer names. To validate our method, we implemented tests that randomly match the names of the developers of our projects. We then manually verified whether the matches were made correctly before performing our analyses. Regarding the strategy used by  stated by Poncin et al.~\cite{MiningRepositories2011}, we considered at least 12 months of commits in a repository as the \textit{``relatively long period of time''} to identify core developers.

Before fitting the OLS model to our data, we verified the correlation among the independent variables of the model (i.e., the 5 CI sub-practices). First, we applied Shapiro-Wilk’s tests \cite{ShapiroTest2021} to verify the normality of our data.  Having observed that none of the independent variables follow a normal distribution, we chose the Kendall non-parametric test \cite{CorrelationTest2021} to measure the correlation between independent variables.

\vspace{1em}

\subsubsection*{RQ2: Which CI sub-practice contribute to higher quality outcomes?}

\hfill \break

\textbf{\textit{\underline{Motivation:}}} 
In RQ2, we aim to better understand which CI sub-practices have a stronger impact on the quality outcomes of a project. Although Vasilescu et al.~\cite{VasilescuQualityProdOutcomes2015} observed that the usage of CI likely holds a relationship with quality outcomes (i.e., number of bug-related IRs), we still do not know whether only certain CI practices have a significant association with quality outcomes. Similar to RQ1, better understanding which CI sub-practices have stronger associations with quality outcomes (i.e., number bug-related IRs~\cite{VasilescuQualityProdOutcomes2015}) may help ups to optimize the usage of CI. 

\textbf{\textit{\underline{Approach:}}} 
Similar to Vasilescu et al.~\cite{VasilescuQualityProdOutcomes2015}, we manually checked our projects to understand how they used tags to indicate the presence of bugs. Similarly to Vasilescu et al.'s work~\cite{VasilescuQualityProdOutcomes2015}, we selected projects that have at least 75\% of their IRs tagged with labels in our dataset. For the sake of understanding how labels were used, we apply this selection criterion in the broader group of 776 relevant projects. We found a total of 121 projects that were manually checked. We found 2 main tags that were used to indicate the presence of bugs (which were not present in the original list \cite{VasilescuQualityProdOutcomes2015}): (i) ``crash'' and (ii) ``regression''. For example, the \textit{mui/material-ui} project\footnote{https://github.com/mui/material-ui/labels?page=5\&sort=name-asc} describes the \textit{``regression''} tag as \textit{``A bug, but worse''}. Thus, we added these 2 new tags to the original list of tags defined by  Vasilescu et al.~\cite{VasilescuQualityProdOutcomes2015}. The resulting list of labels indicating bug-related IRs was:
\textit{ "defect", "error", "bug", "issue", "mistake", "incorrect", "fault" , "flaw", "crash" and "regression"}. We then searched in the group of 90 relevant and active projects (i.e., the final group of projects where we applied our analysis) for these words in IRs' labels. To do so, we performed lowercasing and applied the Porter Stemming Algorithm.\footnote{https://tartarus.org/martin/PorterStemmer/}

Our data for quality outcomes, i.e., number of bug-related IR, is similar in many aspects to the data used in Vasilescu et al.'s work~\cite{VasilescuQualityProdOutcomes2015}: (i) most of our variables are counts (e.g., qty of bugs reported, qty of issues, qty of developers, etc.); (ii) some variables are over-dispersed, i.e., the variance is much larger than the mean; and (iii) some response variables present an excess number of zeros. Because of these characteristics, in RQ2, we applied zero-inflated negative binomial regression models (ZINB) \cite{ZINBR2022}.

To ensure that using ZINB models was an appropriate choice, we first compared poisson regression models with negative binomial regression models. This comparison shows that the residuals are more spread out for the poisson, furthermore, we applied the likelihood ratio test,\footnote{https://www.statisticshowto.com/likelihood-ratio-tests/} which shows a statistically significant difference in favor to negative binomial model. Thus a negative binomial offers a significantly better fit to our data compared to a poisson regression model. Next, we fit a zero-inflated negative binomial model with the ordinary negative binomial model using Vuong’s test\footnote{https://www.rdocumentation.org/packages/pscl/versions/1.5.5/topics/vuong} of a non-nested model. Our test provides evidence of the superiority of a zero-inflated model over an ordinary model (AIC-corrected -3.2992634, p-value $<$ 0.05 ).

\subsubsection*{RQ3: Can consistently applied CI sub-practices be an indicator of process quality?}\label{RQ3}

\hfill \break

\textbf{\textit{\underline{Motivation:}}} Felidré et al. \cite{CITheater2019} observed that not all projects follow all CI sub-practices consistently. For instance, while some projects may have acceptable build duration, they may not perform frequent commits. It is important to know whether projects that perform most of CI sub-practices in a consistent manner are also associated with a higher quality in their development process. Given that it would be challenging to quantitatively measure ``quality of development process'', we perform a qualitative analysis in RQ3.

\begin{figure}[h]
  \centering
  \includegraphics[width=\linewidth]{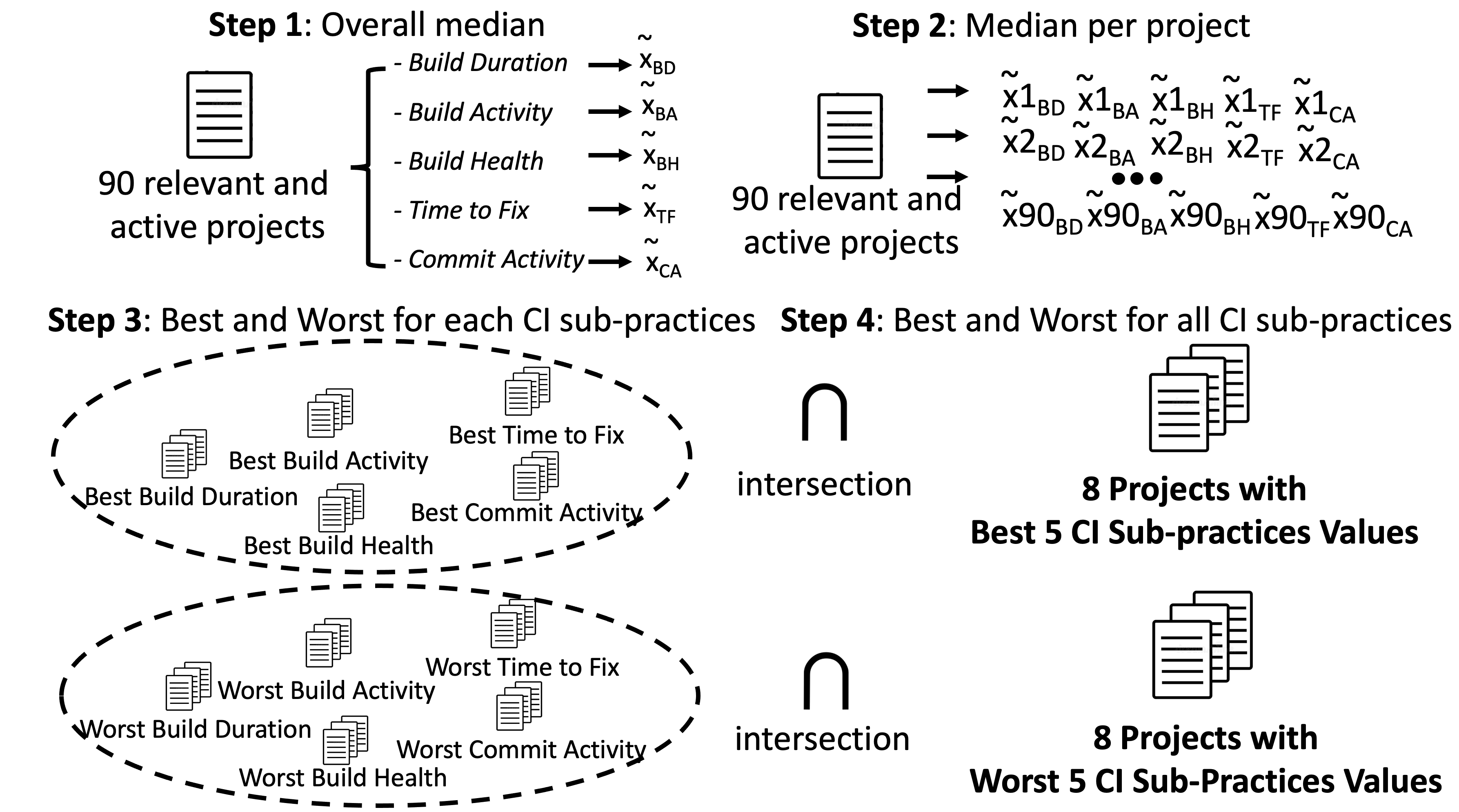}
  \caption{Best and Worst CI sub-practices groups selection}
  \label{fig:bestandworst}
\end{figure}

\textbf{\textit{\underline{Approach:}}} To answer this research question, we separate the projects into 2 groups: (i) projects that follow all the best CI sub-practices analyzed in this study; and (ii) projects that do not effectively follow all the best CI sub-practices. Figure~\ref{fig:bestandworst} shows the steps to generate these two groups of projects.

First, we computed the overall median of each of the 5 CI sub-practices for the 90 relevant and active projects in our dataset. Next, we computed the median of CI sub-practices per project. We then selected projects that had the best and worst values compared to the overall median for each of the 5 CI sub-practices. Lastly, we selected projects in which their medians were all (i) best or (ii) worst the overall median for all 5 CI sub-practices. This segmentation resulted in a group of 8 projects whose medians were above the overall median for all CI sub-practices (the {\em Best CI Practices} group) and another group of equally 8 projects whose medians were below the overall median for all CI sub-practices (the {\em Worst CI Practices} group). We emphasize that the values of the sub-practices have different meanings, depending on the way in which they were calculated. For example, in the case of the Build Duration sub-practice, the best values are lower values. In the case of the Commit Activity sub-practice, the best values are the higher values. These differences were considered when splitting our projects into two groups.

Afterwards, we applied a qualitative approach known as Document Analysis~\cite{DocumentAnalysis2009}. This analysis consists of coding documents into themes. We used the comments of the PRs of the projects as inputs to this analysis. We obtained representative samples of PRs to perform our analyses\cite{SampleSize2021}. The {\em Best CI Practices} group has 1,560 PRs, and the {\em Worst CI Practices} group have 989 PRs. Considering a confidence level of 95\% and a confidence interval of 10\%, we created two representative samples containing 91 PRs and 88 PRs, respectively.

We randomly selected 91 and 88 PRs under the condition that each project would contribute with the same number of PRs inside representative sample. We rounded up the number of PRs and randomly selected 12 PRs for each of the 8 projects within each group. For each of the selected PRs, all comments were collected and save in 2 different files with randomly generated names, which were then used in the thematic analysis. This was important to not identify to which group of projects the PRs belonged. As this is a subjective process, the thematic analysis was performed by two coders--a main coder and a reviewer--to minimize the bias. After the thematic analysis was done, we revealed the data to identify group to which the comments belonged.

\section{Results}

In this section, we present the results of each research question.

\subsection*{RQ1: Which CI sub-practices contribute to higher productivity outcomes?}

Before fitting our regression models, we applied the Kendall test to verify the correlation among the independent variables. The Null Hypothesis is that variables are uncorrelated. If p-value $<$ 0.05, we reject the Null Hypothesis. The $\tau$ value shows how strong the correlation is. \textit{Builds Activity} and \textit{Commit Activity} practices were correlated with $\tau = 0.5109093$ (p-value $<$ 2.2e-16), which is to be expected, since in a CI environment, more commits generate more builds. \textit{Builds Activity} and \textit{Time to Fix a Broken Build} also demonstrated a correlation with $\tau 0.2374206$ (p-value $<$ 2.2e-16). With more builds, projects spend more time to fix them. \textit{Build Health} and \textit{Time to Fix a Broken Build} revealed a negative correlation with $tau -0.4995392 (p-value $<$ 2.2e-16)$, the longer it takes for projects to fix their builds, the less the \textit{Build Health}. Given that our $\tau$ coefficients were not close to 1 (the perfect correlation), we decided to keep all independent variables in our model. 

\begin{figure}[h]
  \centering
  \includegraphics[width=\linewidth]{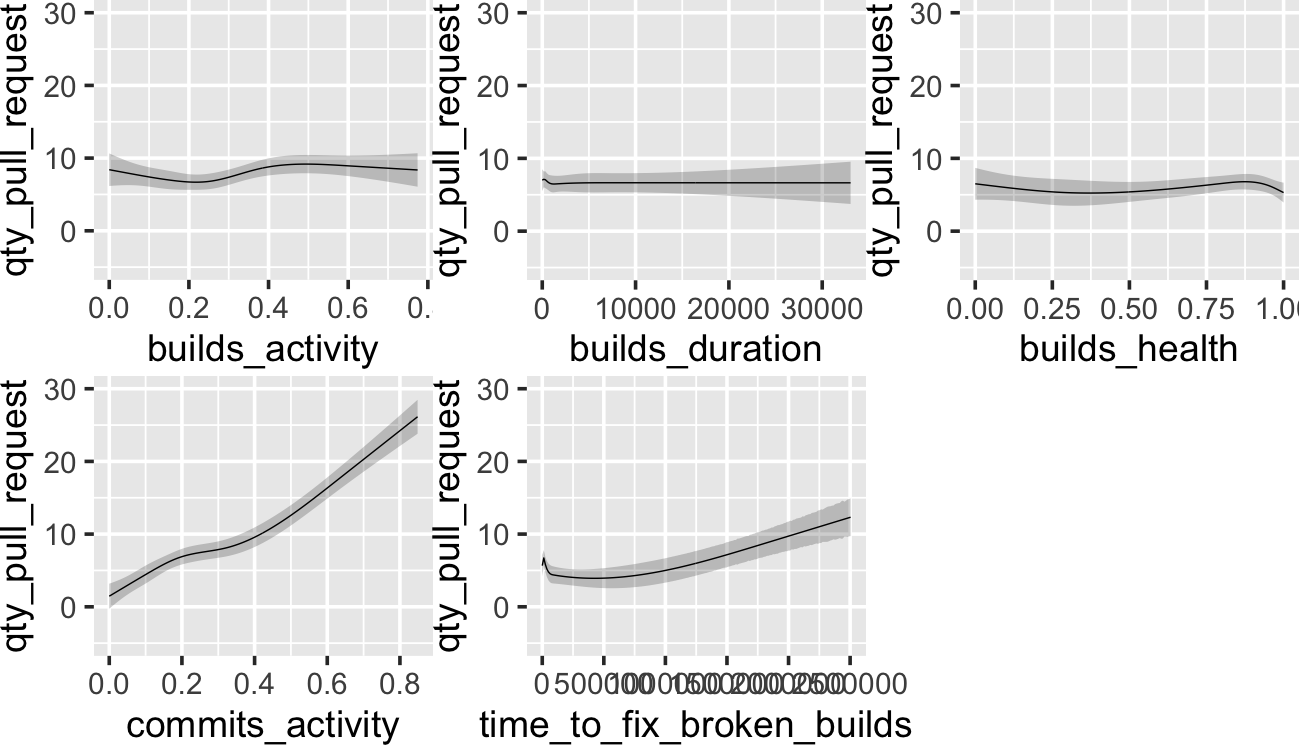}
  \caption{Regression model between the 5 CI sub-practices and the number of merged PRs}
  \label{fig:06_ols_ols_qty_prs}
\end{figure}

Our regression model obtained a R-squared of \textbf{0.353} (R-squared adj. \textbf{0.347}). It is challenging to determine what is a ``good'' R-squared value as it will depend on the goal of the research. 
For example, if the main goal is prediction, R-squared values should be high (e.g., 0.7 to 0.9)~\cite{choi2012predicting}. However, lower R-squared values (e.g., around 0.20) may also provide important insights in psychology or social sciences~\cite{bersani2016association}.
Given that our goal is to better understand associations between CI sub-practices and quality/productivity outcomes, we believe our R-squares value are acceptable. Figure~\ref{fig:06_ols_ols_qty_prs} shows the relationships between the 5 CI sub-practices and the number of merged PRs (as fit by our OLS model).

To verify the strength of the associations between the CI sub-practices and the response variable, we computed Spearman's tests \cite{SpearmanTest2021}. This test measures the strength and direction of associations between two ranked variables by the correlation coefficient $rho$. Our tests reveal that \textit{Commit Activity}, with $\rho = 0.388$, and \textit{Build Activity}, with $\rho = 0.274$, are the two most influential sub-practices when it comes to explain the number of merged PRs. Projects with high \textit{Commit Activity} and \textit{Build Activity} values, tend to be more productive, i.e., more PRs are merged. Although these results are expected, it is interesting to note that sub-practices such as \textit{Time to Fix a Broken Build} do not have much influence on productivity. For example, this may be due to the fact that merges may still occur when the build is broken~\cite{ghaleb2019studying}. Table~\ref{tab:spearman} shows the values of our Spearman's tests.

\begin{table}[htbp]
  \caption{Spearman tests: CI sub-practices vs. merged PRs}
  \label{tab:spearman}
  \begin{center}
  \begin{tabular}{lr}
    \hline
CI sub-practice & rho2 \\ 
\hline
Builds Duration&0.007\\
Builds Activity&\textbf{0.274}\\
Builds Health&0.000\\
Time to Fix Broken Builds&0.014\\
Commits Activity&\textbf{0.388}\\
\hline
\end{tabular}
\end{center}
\end{table}

We fitted another model for the same 5 CI sub-practices in which we divided the number of merged PRs by the number of core and non-core developers in each period of the analysis. We recalculate the regression model to normalize the number of merged PRs by the number of developers and to minimize the effects of the number of developers on the productivity of the projects. This new models obtained lower R-square values (\textbf{0.296} for core developers and \textbf{0.210} for non-core developers) but still revealed significant correlations with CI sub-practices. This result suggests that the number of core and non-core developers may also influence the number of merged PRs but they do not eliminate the potential effect of CI sub-practices on the productivity of outcomes. According to our model, the number of non-core developers has more influence on productivity outcomes than core developers.

\begin{center}
\fbox{
    \begin{minipage}{24em}
    \textbf{
    \textit{Commit Activity} and \textit{Build Activity} are strongly associated with the number of merged Pull Requests. When considering the number of core and non-core developers, the correlations becomes weaker but they are still significant.}
    \end{minipage}
}
\end{center}

\subsection*{RQ2: Which CI sub-practice contribute to higher quality outcomes?}

Table \ref{tab:zeroinflPRNonCore} shows our zero-inflated negative binomial models. We highlight the CI sub-practices that obtain statistical significance.

\begin{table}[htbp]
\caption{ZINB model for bug-related IRs}
\label{tab:zeroinflPRNonCore}
\begin{center}
\begin{tabular}{lllll}
\hline
\multicolumn{5}{l}{Count model coefficients (negbin with log link)} \\ 
\hline
                              & Est. Std. & Error  & z value & Pr(\textgreater{}|z|) \\
(Intercept)                   & -0.1316     & 0.2530 & -0.520  & 0.6027       \\
\textbf{builds  activity}              &  \textbf{1.9846}     & \textbf{0.4163} & \textbf{4.767}    & \textbf{1.87e-06 ***}  \\
\begin{tabular}[c]{@{}l@{}}commits \\ activity\end{tabular}   &  0.7552     & 0.3941 & 1.916    & 0.0553 .     \\
\textbf{builds health}               & \textbf{-0.7466}     & \textbf{0.2710} & \textbf{-2.754}   & \textbf{0.0058 **}    \\
\textbf{time to fix}                & \textbf{-2.4175}     & \textbf{0.5730} & \textbf{-4.219}   & \textbf{2.46e-05 ***}  \\
builds duration              & 3.4774      & 1.8004 & 1.931    & 0.0534 .     \\
Log(theta)                    & -1.5574     & 0.0603 & -25.819  & < 2e-16 ***  \\
\hline
\multicolumn{5}{l}{Zero-inflation model coefficients (binomial with logit link)} \\
\hline
                              & Est. Std. & Error  & z value & Pr(\textgreater{}|z|) \\
(Intercept)                  & 0.1061     &  0.8241  &   0.129  &  0.8975     \\
builds activity             & -10.4734   & 10.0846  &  -1.039  &  0.2990     \\
\textbf{\begin{tabular}[c]{@{}l@{}}commits \\ activity\end{tabular}}            & \textbf{-11.1607}   &  \textbf{5.5541}  &  \textbf{-2.009}  &  \textbf{0.0445} *   \\
builds health               & -11.4912   &  7.0698  &  -1.625  &  0.1041     \\
\textbf{time to fix}                & \textbf{8.0927}     & \textbf{3.9048} & \textbf{2.073}   & \textbf{0.0382 *}  \\
builds duration             & 75.1436    & 43.4784  &  1.728   &  0.0839 .   \\
\hline
\multicolumn{5}{l}{\small Signif. codes:  0 '***' 0.001 '**' 0.01 '*' 0.05 '.' 0.1 ' ' 1} \\
\end{tabular}
\end{center}
\end{table}

The \textit{Build Activity} sub-practice has a significant correlation with the number of bug-related IRs (Table \ref{tab:zeroinflPRNonCore}). This result may be because more builds may be a consequence of the necessity of fixing more bug-related IRs. Another explanation is that more builds may indicate projects that interact more with their end-users who, in turn, report more bug-related IRs. On the other hand, \textit{Build Health} has an inverse correlation with bug-related IRs. The better the \textit{Build Health} the lower the number of  bug-related IRs. This result suggests that projects with less broken builds are less likely to deliver bugs to the end users. Surprisingly, the \textit{Time to Fix a Broken Build} sub-practice has an inverse correlation with bug-related IRs. Projects taking a longer time to fix broken builds may have fewer bug-related IRs.

In our dataset, taking a longer time to fix broken builds is correlated with a reduced \textit{Build Activity} (see Section~RQ1). We conjecture that fewer builds may be associated with a low activity within the projects (e.g., receiving fewer updates or contributions from external users). Consequently, less bug-related IRs could have been reported in projects with a longer \textit{Time to Fix Broken Builds}. In fact, the \textit{Time to Fix Broken Builds} is inversely correlated with \textit{Build Health} (R-squared of 0.2324), meaning that projects that take a longer time to fix a broken build also tend to have more build breakages in general, which may make the project less attractive to users. For example, project stability (e.g., less breakages) may be related to project attractiveness~\cite{yamashita2014magnet}.

Our model also found (in the coefficient parts of the Zero-inflation model) that an increase in Commit Activity is associated with a higher probability of generating bug-related IRs. The Time to Fix Broken Builds increases the odds of being in the group without bug-related IRs. This result reinforces the Count model coefficient part of the ZINB.

\begin{center}
\fbox{
    \begin{minipage}{24em}
    \textbf{
    \textit{Build Activity} has a significant correlation with the the number of bug-related issue reports. Higher values of \textit{Commit Activity} also increase the probability of generating bug-related IRs. Maintaining a good build health seems to decrease the number of bug-related issue reports. However, surprisingly, large values of \textit{Time to Fix a Broken Build} are associated with less bug-related IRs.
    }
    \end{minipage}
}
\end{center}

\subsection*{RQ3: Can consistently applied CI sub-practices be an indicator of process quality?}

\begin{figure}[h]
  \centering
  \includegraphics[width=\linewidth]{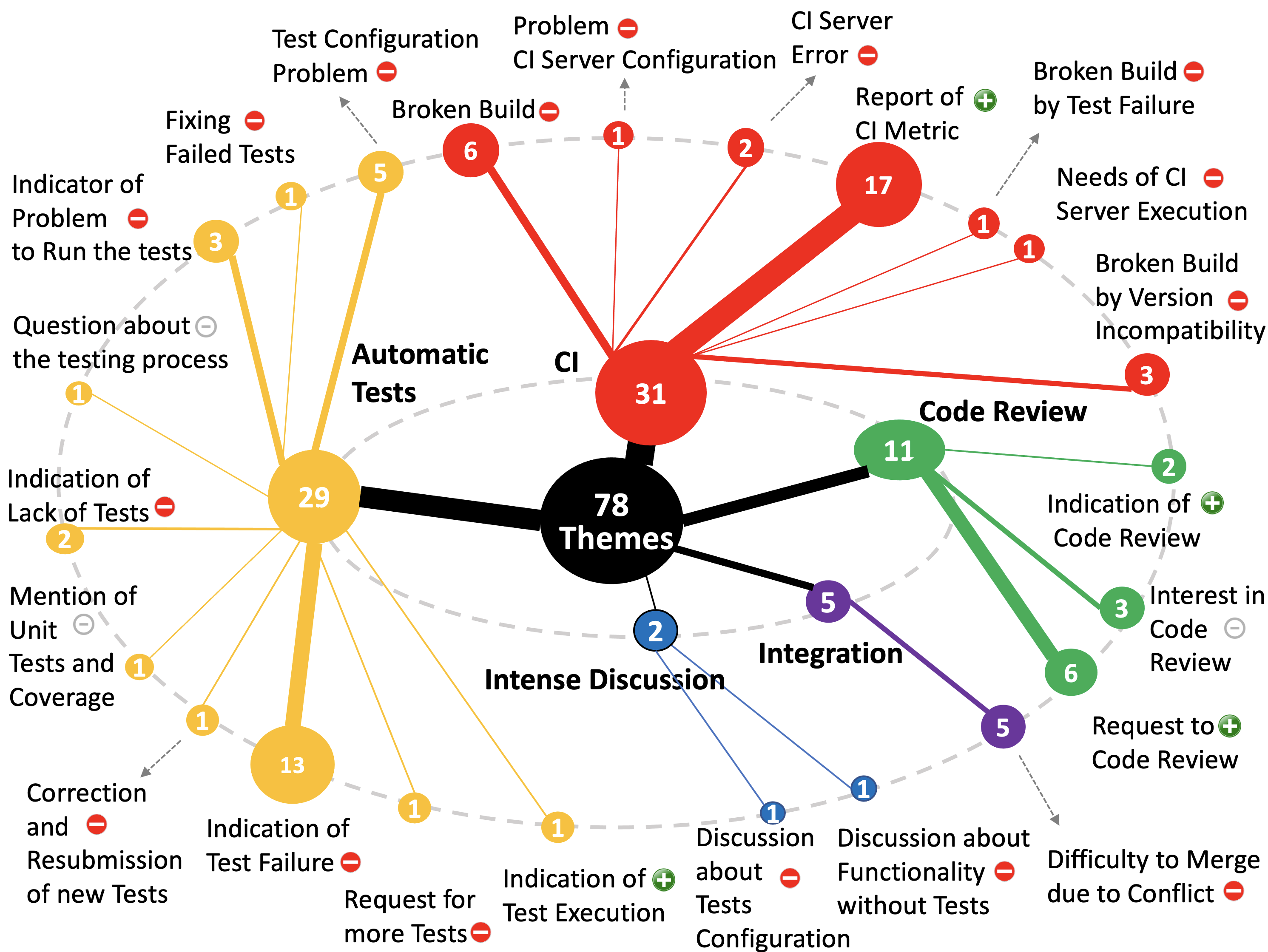}
  \caption{Themes that emerged from the Worst CI projects}
  \label{fig:themes_analysis_worst}
\end{figure}

\begin{figure}[h]
  \centering
  \includegraphics[width=\linewidth]{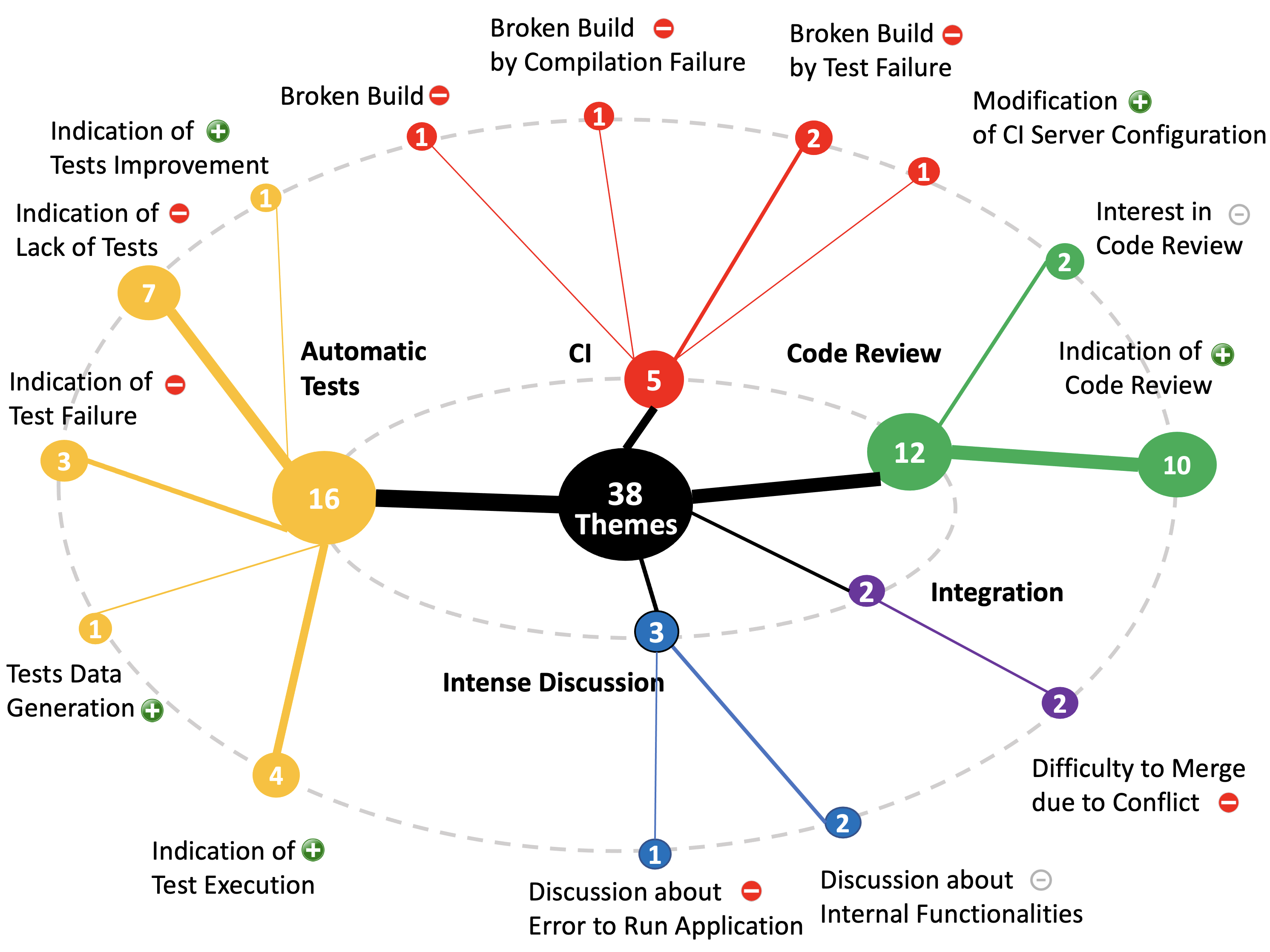}
  \caption{Themes that emerged from the Best CI projects}
  \label{fig:themes_analysis_best}
\end{figure}

Figure~\ref{fig:themes_analysis_worst} illustrates the themes that emerged from our qualitative analysis for the {\em Worst CI projects}, whereas Figure~\ref{fig:themes_analysis_best} shows the themes that emerged from our qualitative analysis for the {\em Best CI projects}. We grouped the themes into 5 CI-related high-level themes. The high-level themes and their definitions are shown in Table~\ref{tab:highlevelthemes}.

\begin{table}[htbp]
  \caption{CI high-level Themes emerged from the Document Analysis}
  \label{tab:highlevelthemes}
  \begin{center}
  \begin{tabular}{ll}
    \hline
CI high-level themes & Description \\ 
\hline
Automatic Tests & \begin{tabular}[c]{@{}l@{}} Denotes evidence of \\ automated tests \end{tabular} \\
Integration & Denotes integration issues \\
Intense Discussion & More than 10 comments in a PR \\
Code Review & \begin{tabular}[c]{@{}l@{}} Indicates that the source code \\ has been reviewed \end{tabular} \\
CI & \begin{tabular}[c]{@{}l@{}} Issues, practices or tools directly \\ related to CI \end{tabular} \\
\hline
\end{tabular}
\end{center}
\end{table}

The number in the center of each node indicates the number of occurrences of each theme. The thickness of the edges and nodes is based on the number of times the theme emerged during the thematic analysis. For example, the \textit{``Discussion about tests configuration''} theme has only one occurrence in the Worst CI projects. The themes were also classified as having a positive ( \includegraphics[width=0.35cm]{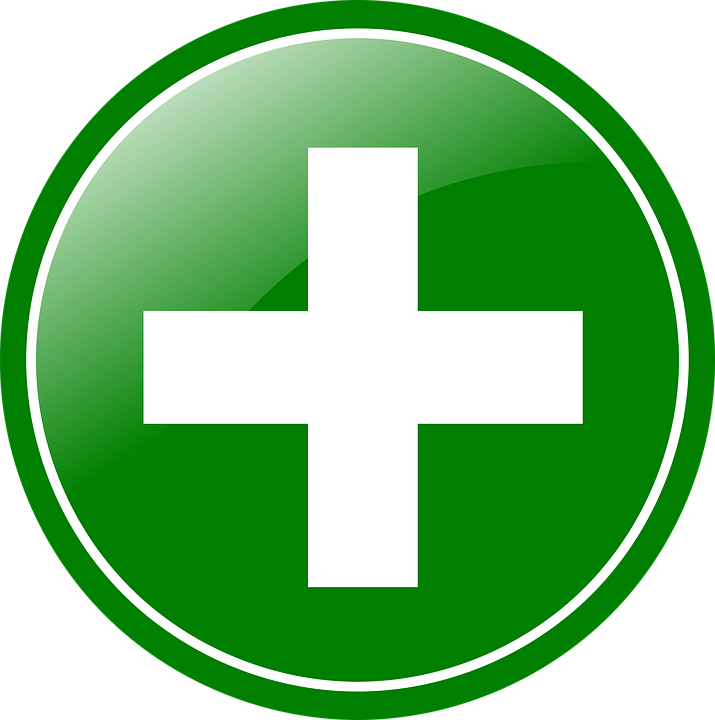} ), neutral ( \includegraphics[width=0.35cm]{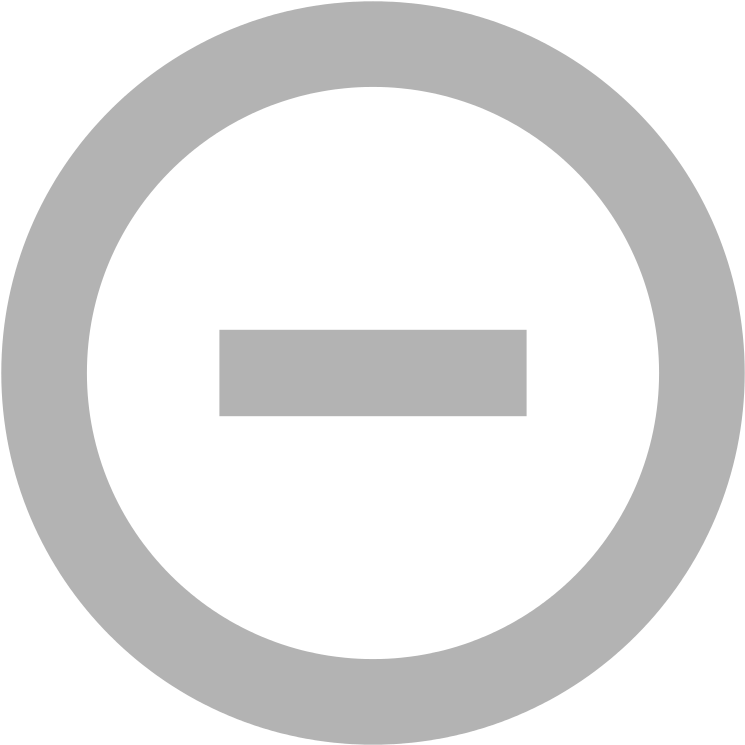} ) or negative ( \includegraphics[width=0.35cm]{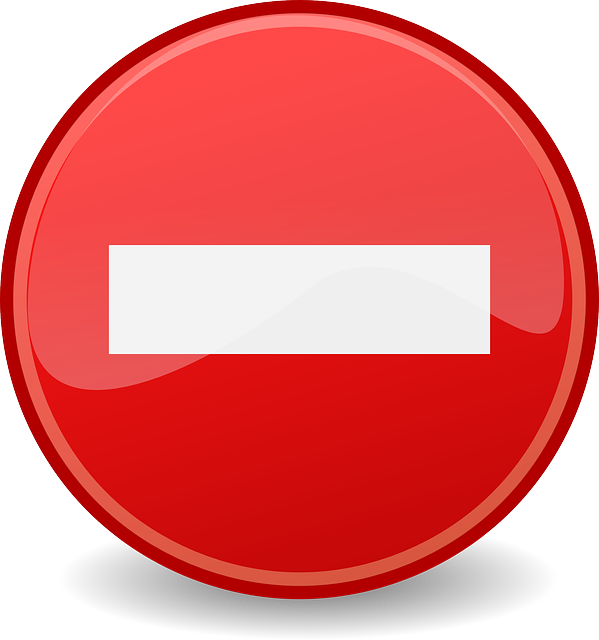} ) meaning in the project under analysis.

As an example of a PR comment that we consider as having a negative meaning, we quote: \textit{``Can you please resolve merging conflicts with changelog  and then I will proceed with merging''}. This comment was associated with the \textbf{``Difficulty to Merge due to Conflict''} theme under the \textbf{``Integration''} high-level theme. Building the software constantly and making frequent commits should reduce conflicts and, consequently, integration effort. 

As an example of a PR comment that we consider as having a positive meaning, we quote: \textit{``fast yet thorough review! I've implemented all of your suggestions''}. This comment was associated with theme \textbf{Indication of Code Review} of the \textbf{Code Review} high-level theme. According to Steve McConnell~\cite{CodeReviewBook04}: \textit{``Code review is one of the most effective ways to find bugs and improve code quality''}, which is a practice recommended in CI. If we find a comment indicating that a code review was accomplished on the project at some point during the implementation of a PR, then we consider that the revised code tends to have fewer bugs~\cite{CodeReviewBook04}. Thus, \textbf{Indication of Code Review} is classified as having a positive meaning. In the opposite case, we quote: \textit{``Thanks for looking into the test failure!''} This comment was associated with theme \textbf{Indication of Test Failure} of the \textbf{Automatic Tests} high-level theme. According to the grey literature: \cite{InfoWorld2007}: \textit{``Developers must write automated builds with tests that pass 100\% of the time, and not get or commit broken code from/to the version control repository''}. If a PR comment indicates that "tests failed" it has a negative aspect for the project. Thus, we classified it as having a negative meaning. We explain our themes more thoroughly below.

{\bfseries Automated Tests.} We obtained 29 citations related to {\em Automated Tests} in the Worst CI projects, of which 13 citations were related to the \textbf{Indication of Test Failure (negative)} theme. As examples of comments indicating test failures in the Worst CI projects, we quote the following comments:  \textit{``This works great in my tests with Postman against my rails app  but the specs I added fail in a mysterious way.''} In another PR, a developers comments: \textit{``On the tests: The first two fail because `PORO::Employee\#positions` will return an empty array by default''}. We also observed occurrences of specific tests failures in comments such as:  \textit{``tests that are failing are being run with Python 2.7''}. Regarding the Best CI projects, we only found 3 citations related to the \textbf{Indication of Test Failure} theme. As an example, we quote: \textit{``Init test for hash type 6211. Could not load test module: m06211.pm''} and  \textit{``the test fail is unrelated to your changes.''}.

{\bfseries Code Review.} We found only two citations of the {\bfseries Indication of Code Review (positive)} theme in the Worst CI projects, whereas we found 10 citations of the same theme in the Best CI projects. As examples, we quote the following excerpts for the Worst CI projects: \textit{``I couldn't figure it out. Otherwise it is ready for review''}. For the Best CI projects we quote:  \textit{``I started reviewing some of the code ''}, \textit{``I see more changes that are needed  but it looks already very good. Good job. ''}, \textit{``another question: did you @xxxxxx  try with much longer salt size''}, and \textit{``Processed your review comments...''}. 

{\bfseries Integration.} We found five citations related to the {\bfseries Difficulty to Merge due to Conflict (negative)} theme in the Worst CI projects, whereas we found two citations of the same theme in the Best CI projects. As examples, we quote the following excerpts for the Worst CI projects: \textit{``I rebased this branch on master resolved conflicts and merged  then force pushed''}, \textit{``Didn't need to bump the examples version  but it doesn't really matter. Another conflict popped up however.''}. For the Best CI projects we quote:  \textit{``Looks good now  but I can't merge because of docs/credits.txt conflict.''}.

{\bfseries Intense Discussion.} We found two main themes in the Worst CI projects: {\bfseries Discussion about
tests configuration (negative)} and {\bfseries Discussion on functionalities without tests (negative)}. On the other hand, we found one negative theme in the Best CI projects: {\bfseries Discussion on errors when running the application}. As examples, we quote the following excerpts for the Worst CI projects: \textit{``This behavior has no tests and is not documented.''} and \textit{``Let's say hypothetically I moved the tests out of the package. On a scale from -1 to +1 how annoyed would you be to re-sync this again?.''}. For the Best CI projects we quote: \textit{``I believe the line `texexpand: include titlepag.tex failed.` could be the cause for the index.html being malformed.''}

{\bfseries Continuous Integration (CI).} The Worst CI projects obtained 7 different themes directly related to the \textbf{CI} high-level theme, whereas the Best CI projects obtained only four \textbf{CI} themes. We thought this result was counter-intuitive. However, from the 7 different themes in the Worst CI projects, 6 of them were negative. Some examples of negative comments were: \textit{``The build is failing -- would you take a look to see if you can get it working''}, \textit{``Travis CI is being derpy again so whenever that decides to work.''},   \textit{``Removing the archive makes our builds reproducible without setting that environment variable.''}, \textit{``Travis CI failed only ruby 2.2.10.  I will rebuild Travis CI.''}, \textit{``I see that this is failing to pass the tests in Travis - I'll have a go at fixing that up now''},  \textit{``Closed and reopened to wake up Travis CI.''}. Regarding the Best CI Projects, from the four themes that we found, three themes were negative and one theme was positive. The positive theme was the \textbf{Modification of CI Server Configuration} theme: \textit{``I modified the Travis configuration anyways to make builds ~30 seconds faster.''}

The theme \textbf{Reporting of a CI Metric} has a greater occurrence in the Worst CI projects and no occurrence at all in the Best CI projects. Indeed, this happened because the \textit{``graphite-project/graphite-web''} project (classified under the Worst CI projects) used the \textsc{CodeCov} tool\footnote{https://about.codecov.io/product/feature/pull-request-comments/}, which automatically reports coverage values in the comments of each PR. Despite being a positive practice, unfortunately, the usage of such a tool does match the behavior of the project, i.e., the project obtained lower values within all CI sub-practices. 

In general, the Worst CI projects have more citations of CI-related themes (78). However, most of these comments (60\%) were classified as having a negative meaning. If we do not consider the {\bfseries Reporting of a CI Metric} theme, the percentage would rise to (77\%). The Best CI projects obtained 38 CI-related themes with 17 (44\%) of the themes classified as having a negative meaning. For all CI high-level themes, the results were favorable for Best CI projects. Therefore, our data and results suggest that consistently employing CI sub-practices may indeed be associated with an improved development process.

\begin{center}
\fbox{
    \begin{minipage}{24em}
    \textbf{
    Our results reveal that projects with best CI sub-practices values have fewer citations related to the CI high-level themes and these citations have a more positive meaning. It suggests that higher values in the 5 CI sub-practices indeed can indicate a better quality in the development process.
    }
    \end{minipage}
}
\end{center}

\section{Discussion}
The main goal of our work was to highlight the importance of studying CI by means of its sub-practices. We were inspired by Felidré et al.'s work on the {\em CI Theater} phenomenon~\cite{CITheater2019}, which revealed that many projects do not follow CI practices. Therefore, instead of considering the usage of a CI server as the sole criterion to assume that projects use CI (which is a problem also highlighted by Soares et al.'s work~\cite{EliezioRevisaoSistematica2022}), we investigated the potential impact of CI on quality and productivity outcomes by means of 5 CI sub-practices. 

All analyzed projects in our study adopt a CI service to automate the build and test execution process. However, only the most active projects --- the projects making frequent commits and builds --- are more productive. Furthermore, the increase in commit and build frequency does not come at the cost of quality, meaning that frequent commits and builds are more effective to generate more delivery value into the project. Additionally, CI can make developers more confident as they witness automated builds and tests providing continuous feedback. Thus pull requests are likely only created when  they are in a near-approval state.

On the other hand, more active projects (in terms of builds and commits) tend to have more bug-related issue reports open. This can be an indication that such projects have a higher demand and receive more requests from external users or internally from the developers. Our results suggests that maintaining a good build health has a correlation with code quality. A better build health is associated with more stable versions of the project, generating fewer errors for users and consequently fewer bug-related issue reports.

Our work also found that projects with the best values for CI sub-practices generated fewer CI-related themes when we analyzed their development processes. In fact, the CI-related themes found in projects adhering to the sub-practices usually have a more positive connotation within the project. It reveals that projects that follow CI best sub-practices can be a reflection of a shared understanding between developers on how CI must be used, allowing developers to focus more on the business rules instead of discussing how CI should be used. Perhaps, if a project can document its ``Continuous Integration Principles'' it would make it easier for new contributors to understand the philosophy of the project when it comes to CI usage.  

Next we discuss the implications of our work for researchers and practitioners.

\textbf{Implications for researchers: }
Our study shows that we need to consider, measure and evaluate the different CI sub-practices to better determine that CI is being used in a project. Especially in RQ3, when we segment our projects into different groups (i.e., Best and Worst CI practices), we have evidence that the developers' experience with CI is quite different. Future research should consider evaluating several CI sub-practices before evaluating the potential influence of CI adoption on software development aspects.

\textbf{Implications for practitioners: } 
The results of RQ1 and RQ2 indicate that certain benefits of CI (e.g., more merged PRs) are not equally shared by all of the CI sub-practices. For instance, our models in RQ1 reveal that practices such as \textit{Build Health} and \textit{Build Duration} do not have much influence on the number of merged PRs. Our results suggest that, instead of fully adopting CI from the beginning, a staged adoption (e.g., only certain practices) may be more wise given the goals of the project. For example, starting with frequent commits and builds Moreover, given that {\em culture} is an important factor for the successful adoption of new methodologies~\cite{gupta2019relationships}, the staged CI adoption can be a CI-enabler. With respect to quality, our recommendation is to strive to maintain a sound build health, i.e., to generate fewer broken builds. Build health obtained a significant and inverse correlation with the number of bug-related issue reports.

\section{Threats to Validity}

In this section, we discuss the threats to the validity of our study.

\textbf{Construct Validity Threats:} Although \textsc{Travis-CI} is widely used in the context of \textsc{GitHub} open-source projects, existing projects may use other CI services or servers, which might have led us to discard relevant projects.

With respect to the analysis regarding core developers, there is no unique identifier to compare \textsc{GitHub} developers (as mentioned by Zhao et al. \cite{VasilescuImpactCI2017}). Due to this issue, we applied a heuristic based on the comparison of developer names to collect the data associated with a specific core or non-core developer. This heuristic may not be the most accurate, generating false negatives or positives in our data. We used sub-samples of the investigated projects to assess the quality of the extracted information related to developers.

\textbf{Internal Validity Threats:} We collected data regarding productivity and quality outcomes and CI sub-practices using implementation strategies reported by previous work. The  productivity and quality outcomes were quantified in our study by means of merged PRs and bug-related closed issue reports, respectively, inspired by Vasilescu et al.'s work~\cite{VasilescuQualityProdOutcomes2015}. However, these measures only represent a specific perspective of productivity and quality. Other existing perspectives (e.g., effort) could be explored in future work. Our study also explored a total of 5 sub-practices related to CI. Although they are expressive, this set could be extended with the analysis of other existing sub-practices such as code coverage, for example. In fact, we did not include the code coverage analysis because we only found a few \textsc{GitHub} projects maintaining a history of this metric in existing public repositories. Felidré et al.'s work~\cite{CITheater2019} found the same obstacle when analyzing existing CI practices on \textsc{GitHub} projects.  

In the qualitative analysis of RQ3, we acknowledge the potential for bias from the authors' subjective interpretations of PR comments related to CI sub-practices. We mitigate this threat through the peer review of the codes and themes extracted using document analysis. As mentioned in Section 2, we also omitted the groups of projects of the PR comments that were analyzed to avoid bias related to some specific group (i.e., \textit{The Best and Worst CI Practices groups}.)

\textbf{External Validity Threats: } All analyzed open-source projects are hosted on the \textsc{GitHub} platform and use  \textsc{Travis-CI} as their CI service. We acknowledge that our results are restricted to the context of the analyzed projects. Additional replication studies are necessary in future work to generalize the results for projects of different nature. We made our dataset and results available to allow future replications of our study.\footnote{https://zenodo.org/record/6513155}

\section{Related Work}

In this section, we situate our study with respect to previous works that investigate the relationship between CI sub-practices and the productivity and quality of open-source projects.

Vasilescu et al. \cite{VasilescuQualityProdOutcomes2015} developed a study to discern the effects of CI adoption in quality and productivity outcomes. They collected a dataset of 246 GitHub projects which at some point in their history added the Travis CI to the development process. They found that after adopting the Travis CI, teams are significantly more effective at merging pull requests submitted by core members. They also report that core developers in teams using CI are able to discover significantly more bugs than in teams not using CI. Similar to Vasilescu's et al. work, our study investigated the CI impact on the productivity and quality of popular open-source projects by using the merged PR and closed bug-related issue outcomes, respectively. On the other hand, we analyzed the influence of different CI sub-practices on these outcomes. Our study found similar results for: (i) productivity - in which the \textit{Commit Activity and Build Activity} CI sub-practices have impacted positively in the number of merged PRs; and (ii) quality - where the \textit{Build Activity} CI sub-practice is correlated with an increase in the number of closed bug-related issues.

Bernardo et al. \cite{JoaoHelis2018} analyzed 162,653 pull requests of 87 GitHub projects that are implemented in 5 different programming languages to evaluate the impact of adopting CI on the time to deliver merged PRs. Their work also considers the beginning of the usage of Travis online service as the starting point for the CI adoption. The study reports  a large increase in the number of submitted, merged and delivered PRs per release after the CI adoption. Similar to Bernardo et al. work \cite{JoaoHelis2018}, our work also found an increase in the number of merged PRs but correlated with the \textit{Commit Activity and Build Activity} CI sub-practices. In addition to that, we investigated the improvement of the quality of the project in terms of closed bug-related issues.

Felidré et al. \cite{CITheater2019} analyzed 1,270 open-source projects that use Travis CI and quantitatively studied the behavior of the adoption of existing CI sub-practices. They found that: (i) 60\% of the projects have infrequent commits; (ii) 85\% of the projects have at least one broken build that take a long time to be fixed; and (ii) most of projects have a build that executes with more than 10 minutes. Similar to Felidré et al. \cite{CITheater2019}, we found indication that many Travis CI projects do not pay attention to the behavior of CI sub-practices during their development and evolution. However, the main goal of our work was assessing the benefits of these practices for the productivity and quality of projects. As our results showed, tracking these sub-practices is important to merge more PRs, generate fewer errors and face fewer problems during the development process.

\section{Conclusion}

We conducted an exploratory quantitative and qualitative study investigating how Continuous Integration practices may influence the productivity and quality outcomes of open-source projects. We analyzed a set of 90 relevant and active open-source projects for a period of 2 years.

We applied linear regression models to study potential associations between five CI sub-practices and the number of merged PRs (productivity). In addition, we studied the potential associations between the five CI sub-practices and the number of bug-related issue reports (quality).

Our findings revealed a positive correlation between the {\em Commit Activity} and {\em Build Activity} CI sub-practices and the increase in the number of merged pull requests. We also observed that {\em Build Activity} has a significant correlation with the number of bug-related issue reports. Moreover, a sound {\em Build Health} is associated with a decrease in the number of bug-related issue reports. Lastly, larger values of {\em Time to Fix a Broken Build} are associated with less bug-related issues. 

To complement our quantitative analysis, we performed a qualitative Document Analysis to identify CI-related themes in the comments of pull requests. Our analysis suggest that higher values in the five analyzed CI sub-practices can indicate a better quality in the development process.

\begin{acks}
This work is partially supported by INES (www.ines.org.br), CNPq grant 465614/2014-0, CAPES grant 88887.136410/2017-00, and \\*FACEPE grants APQ-0399-1.03/17 and PRONEX APQ/0388-1.03/14.

\end{acks}

\vspace{1em}

\balance
\bibliographystyle{ACM-Reference-Format}
\bibliography{bibliography}


\begin{thebibliography}{34}


\ifx \showCODEN    \undefined \def \showCODEN     #1{\unskip}     \fi
\ifx \showDOI      \undefined \def \showDOI       #1{#1}\fi
\ifx \showISBNx    \undefined \def \showISBNx     #1{\unskip}     \fi
\ifx \showISBNxiii \undefined \def \showISBNxiii  #1{\unskip}     \fi
\ifx \showISSN     \undefined \def \showISSN      #1{\unskip}     \fi
\ifx \showLCCN     \undefined \def \showLCCN      #1{\unskip}     \fi
\ifx \shownote     \undefined \def \shownote      #1{#1}          \fi
\ifx \showarticletitle \undefined \def \showarticletitle #1{#1}   \fi
\ifx \showURL      \undefined \def \showURL       {\relax}        \fi
\providecommand\bibfield[2]{#2}
\providecommand\bibinfo[2]{#2}
\providecommand\natexlab[1]{#1}
\providecommand\showeprint[2][]{arXiv:#2}

\bibitem[Allamanis et~al\mbox{.}(2016)]%
        {ConvolutionalAttentionNetwork2017}
\bibfield{author}{\bibinfo{person}{Miltiadis Allamanis}, \bibinfo{person}{Hao
  Peng}, {and} \bibinfo{person}{Charles Sutton}.}
  \bibinfo{year}{2016}\natexlab{}.
\newblock \showarticletitle{A Convolutional Attention Network for Extreme
  Summarization of Source Code}.
\newblock \bibinfo{journal}{\emph{CoRR}}  \bibinfo{volume}{abs/1602.03001}
  (\bibinfo{year}{2016}).
\newblock
\showeprint[arXiv]{1602.03001}
\urldef\tempurl%
\url{http://arxiv.org/abs/1602.03001}
\showURL{%
\tempurl}


\bibitem[Bernardo et~al\mbox{.}(2018)]%
        {JoaoHelis2018}
\bibfield{author}{\bibinfo{person}{Jo\~{a}o~Helis Bernardo},
  \bibinfo{person}{Daniel~Alencar da Costa}, {and} \bibinfo{person}{Uir\'{a}
  Kulesza}.} \bibinfo{year}{2018}\natexlab{}.
\newblock \showarticletitle{Studying the Impact of Adopting Continuous
  Integration on the Delivery Time of Pull Requests}. In
  \bibinfo{booktitle}{\emph{Proceedings of the 15th International Conference on
  Mining Software Repositories}}. \bibinfo{publisher}{Association for Computing
  Machinery}, \bibinfo{address}{New York, NY, USA}, \bibinfo{pages}{131–141}.
\newblock
\showISBNx{9781450357166}
\urldef\tempurl%
\url{https://doi.org/10.1145/3196398.3196421}
\showDOI{\tempurl}


\bibitem[Bersani et~al\mbox{.}(2016)]%
        {bersani2016association}
\bibfield{author}{\bibinfo{person}{Francesco~S Bersani},
  \bibinfo{person}{Daniel Lindqvist}, \bibinfo{person}{Synthia~H Mellon},
  \bibinfo{person}{Elissa~S Epel}, \bibinfo{person}{Rachel Yehuda},
  \bibinfo{person}{Janine Flory}, \bibinfo{person}{Clare Henn-Hasse},
  \bibinfo{person}{Linda~M Bierer}, \bibinfo{person}{Iouri Makotkine},
  \bibinfo{person}{Duna Abu-Amara}, {et~al\mbox{.}}}
  \bibinfo{year}{2016}\natexlab{}.
\newblock \showarticletitle{Association of dimensional psychological health
  measures with telomere length in male war veterans}.
\newblock \bibinfo{journal}{\emph{Journal of affective disorders}}
  \bibinfo{volume}{190} (\bibinfo{year}{2016}), \bibinfo{pages}{537--542}.
\newblock


\bibitem[Bowen(2009)]%
        {DocumentAnalysis2009}
\bibfield{author}{\bibinfo{person}{Glenn Bowen}.}
  \bibinfo{year}{2009}\natexlab{}.
\newblock \showarticletitle{Document Analysis as a Qualitative Research
  Method}.
\newblock \bibinfo{journal}{\emph{Qualitative Research Journal}}
  \bibinfo{volume}{9} (\bibinfo{date}{08} \bibinfo{year}{2009}),
  \bibinfo{pages}{27--40}.
\newblock
\urldef\tempurl%
\url{https://doi.org/10.3316/QRJ0902027}
\showDOI{\tempurl}


\bibitem[Choi and Varian(2012)]%
        {choi2012predicting}
\bibfield{author}{\bibinfo{person}{Hyunyoung Choi} {and} \bibinfo{person}{Hal
  Varian}.} \bibinfo{year}{2012}\natexlab{}.
\newblock \showarticletitle{Predicting the present with Google Trends}.
\newblock \bibinfo{journal}{\emph{Economic record}}  \bibinfo{volume}{88}
  (\bibinfo{year}{2012}), \bibinfo{pages}{2--9}.
\newblock


\bibitem[Cuelogic(2017)]%
        {LevenshteinDistance2017}
\bibfield{author}{\bibinfo{person}{Cuelogic}.} \bibinfo{year}{2017}\natexlab{}.
\newblock \bibinfo{booktitle}{\emph{The Levenshtein Algorithm}}.
\newblock
\urldef\tempurl%
\url{https://www.cuelogic.com/blog/the-levenshtein-algorithm}
\showURL{%
Retrieved december 02, 2021 from \tempurl}


\bibitem[de~Oliveira(2017)]%
        {MarcosOliverira2017}
\bibfield{author}{\bibinfo{person}{Marcos~C\'{e}sar de Oliveira}.}
  \bibinfo{year}{2017}\natexlab{}.
\newblock \showarticletitle{DRACO: Discovering Refactorings That Improve
  Architecture Using Fine-Grained Co-Change Dependencies}. In
  \bibinfo{booktitle}{\emph{Proceedings of the 2017 11th Joint Meeting on
  Foundations of Software Engineering}} (Paderborn, Germany)
  \emph{(\bibinfo{series}{ESEC/FSE 2017})}. \bibinfo{publisher}{Association for
  Computing Machinery}, \bibinfo{address}{New York, NY, USA},
  \bibinfo{pages}{1018–1021}.
\newblock
\showISBNx{9781450351058}
\urldef\tempurl%
\url{https://doi.org/10.1145/3106237.3119872}
\showDOI{\tempurl}


\bibitem[{de Oliveira} et~al\mbox{.}(2019)]%
        {DEOLIVEIRA2019110420}
\bibfield{author}{\bibinfo{person}{Marcos~César {de Oliveira}},
  \bibinfo{person}{Davi Freitas}, \bibinfo{person}{Rodrigo Bonifácio},
  \bibinfo{person}{Gustavo Pinto}, {and} \bibinfo{person}{David Lo}.}
  \bibinfo{year}{2019}\natexlab{}.
\newblock \showarticletitle{Finding needles in a haystack: Leveraging co-change
  dependencies to recommend refactorings}.
\newblock \bibinfo{journal}{\emph{Journal of Systems and Software}}
  \bibinfo{volume}{158} (\bibinfo{year}{2019}), \bibinfo{pages}{110420}.
\newblock
\showISSN{0164-1212}
\urldef\tempurl%
\url{https://doi.org/10.1016/j.jss.2019.110420}
\showDOI{\tempurl}


\bibitem[Duvall et~al\mbox{.}(2007)]%
        {Duvall2007}
\bibfield{author}{\bibinfo{person}{Paul Duvall}, \bibinfo{person}{Stephen~M.
  Matyas}, {and} \bibinfo{person}{Andrew Glover}.}
  \bibinfo{year}{2007}\natexlab{}.
\newblock \bibinfo{booktitle}{\emph{Continuous Integration: Improving Software
  Quality and Reducing Risk (The Addison-Wesley Signature Series)}}.
\newblock \bibinfo{publisher}{Addison-Wesley Professional}.
\newblock
\showISBNx{0321336380}


\bibitem[Felidré et~al\mbox{.}(2019)]%
        {CITheater2019}
\bibfield{author}{\bibinfo{person}{Wagner Felidré}, \bibinfo{person}{Leonardo
  Furtado}, \bibinfo{person}{Daniel A.~da Costa}, \bibinfo{person}{Bruno
  Cartaxo}, {and} \bibinfo{person}{Gustavo Pinto}.}
  \bibinfo{year}{2019}\natexlab{}.
\newblock \showarticletitle{Continuous Integration Theater}. In
  \bibinfo{booktitle}{\emph{2019 ACM/IEEE International Symposium on Empirical
  Software Engineering and Measurement (ESEM)}}. \bibinfo{pages}{1--10}.
\newblock
\urldef\tempurl%
\url{https://doi.org/10.1109/ESEM.2019.8870152}
\showDOI{\tempurl}


\bibitem[Fowler(2006)]%
        {FowlerCI2006}
\bibfield{author}{\bibinfo{person}{Martin Fowler}.}
  \bibinfo{year}{2006}\natexlab{}.
\newblock \bibinfo{booktitle}{\emph{Continuous Integration}}.
\newblock
\urldef\tempurl%
\url{https://martinfowler.com/articles/continuousIntegration.html}
\showURL{%
Retrieved november 24, 2021 from \tempurl}


\bibitem[GeeksforGeeks(2021)]%
        {JaroWinklerSimilarity2020}
\bibfield{author}{\bibinfo{person}{GeeksforGeeks}.}
  \bibinfo{year}{2021}\natexlab{}.
\newblock \bibinfo{booktitle}{\emph{Jaro and Jaro-Winkler similarity}}.
\newblock
\urldef\tempurl%
\url{https://www.geeksforgeeks.org/jaro-and-jaro-winkler-similarity/}
\showURL{%
Retrieved december 02, 2021 from \tempurl}


\bibitem[Ghaleb et~al\mbox{.}(2019a)]%
        {DanielBuildDuration2019}
\bibfield{author}{\bibinfo{person}{Taher Ghaleb}, \bibinfo{person}{Daniel
  Costa}, {and} \bibinfo{person}{Ying Zou}.} \bibinfo{year}{2019}\natexlab{a}.
\newblock \showarticletitle{An Empirical Study of the Long Duration of
  Continuous Integration Builds}.
\newblock \bibinfo{journal}{\emph{Empirical Software Engineering}}
  \bibinfo{volume}{24} (\bibinfo{date}{08} \bibinfo{year}{2019}).
\newblock
\urldef\tempurl%
\url{https://doi.org/10.1007/s10664-019-09695-9}
\showDOI{\tempurl}


\bibitem[Ghaleb et~al\mbox{.}(2019b)]%
        {ghaleb2019studying}
\bibfield{author}{\bibinfo{person}{Taher~Ahmed Ghaleb},
  \bibinfo{person}{Daniel~Alencar da Costa}, \bibinfo{person}{Ying Zou}, {and}
  \bibinfo{person}{Ahmed~E Hassan}.} \bibinfo{year}{2019}\natexlab{b}.
\newblock \showarticletitle{Studying the impact of noises in build breakage
  data}.
\newblock \bibinfo{journal}{\emph{IEEE Transactions on Software Engineering}}
  (\bibinfo{year}{2019}).
\newblock


\bibitem[Group(2022)]%
        {ZINBR2022}
\bibfield{author}{\bibinfo{person}{UCLA: Statistical~Consulting Group}.}
  \bibinfo{year}{2022}\natexlab{}.
\newblock \bibinfo{booktitle}{\emph{ZERO-INFLATED NEGATIVE BINOMIAL REGRESSION
  | R DATA ANALYSIS EXAMPLES}}.
\newblock
\urldef\tempurl%
\url{https://stats.oarc.ucla.edu/r/dae/zinb/}
\showURL{%
Retrieved may 02, 2022 from \tempurl}


\bibitem[Gupta et~al\mbox{.}(2019)]%
        {gupta2019relationships}
\bibfield{author}{\bibinfo{person}{Manjul Gupta}, \bibinfo{person}{Joey~F
  George}, {and} \bibinfo{person}{Weidong Xia}.}
  \bibinfo{year}{2019}\natexlab{}.
\newblock \showarticletitle{Relationships between IT department culture and
  agile software development practices: An empirical investigation}.
\newblock \bibinfo{journal}{\emph{International Journal of Information
  Management}}  \bibinfo{volume}{44} (\bibinfo{year}{2019}),
  \bibinfo{pages}{13--24}.
\newblock


\bibitem[Harrell(2022)]%
        {OLS2022}
\bibfield{author}{\bibinfo{person}{Frank Harrell}.}
  \bibinfo{year}{2022}\natexlab{}.
\newblock \bibinfo{booktitle}{\emph{ols: Linear Model Estimation Using Ordinary
  Least Squares}}.
\newblock
\urldef\tempurl%
\url{https://rdrr.io/cran/rms/man/ols.html}
\showURL{%
Retrieved may 02, 2022 from \tempurl}


\bibitem[Hilton et~al\mbox{.}(2017)]%
        {TradeOffsCI2017}
\bibfield{author}{\bibinfo{person}{Michael Hilton}, \bibinfo{person}{Nicholas
  Nelson}, \bibinfo{person}{Timothy Tunnell}, \bibinfo{person}{Darko Marinov},
  {and} \bibinfo{person}{Danny Dig}.} \bibinfo{year}{2017}\natexlab{}.
\newblock \showarticletitle{Trade-Offs in Continuous Integration: Assurance,
  Security, and Flexibility}. \bibinfo{publisher}{Association for Computing
  Machinery}, \bibinfo{address}{New York, NY, USA}.
\newblock
\showISBNx{9781450351058}
\urldef\tempurl%
\url{https://doi.org/10.1145/3106237.3106270}
\showDOI{\tempurl}


\bibitem[KIBUACHA(2021)]%
        {SampleSize2021}
\bibfield{author}{\bibinfo{person}{FRANKLINE KIBUACHA}.}
  \bibinfo{year}{2021}\natexlab{}.
\newblock \showarticletitle{How to Determine Sample Size for a Research Study}.
\newblock  (\bibinfo{year}{2021}).
\newblock
\urldef\tempurl%
\url{https://www.geopoll.com/blog/sample-size-research/}
\showURL{%
\tempurl}


\bibitem[Laukkanen et~al\mbox{.}(2015)]%
        {StakeholderPerceptions2015}
\bibfield{author}{\bibinfo{person}{Eero Laukkanen}, \bibinfo{person}{Maria
  Paasivaara}, {and} \bibinfo{person}{Teemu Arvonen}.}
  \bibinfo{year}{2015}\natexlab{}.
\newblock \showarticletitle{Stakeholder Perceptions of the Adoption of
  Continuous Integration -- A Case Study}. In \bibinfo{booktitle}{\emph{2015
  Agile Conference}}. \bibinfo{pages}{11--20}.
\newblock
\urldef\tempurl%
\url{https://doi.org/10.1109/Agile.2015.15}
\showDOI{\tempurl}


\bibitem[McConnell(2004)]%
        {CodeReviewBook04}
\bibfield{author}{\bibinfo{person}{Steve McConnell}.}
  \bibinfo{year}{2004}\natexlab{}.
\newblock \bibinfo{booktitle}{\emph{Code Complete: A Practical Handbook of
  Software Construction} (\bibinfo{edition}{2} ed.)}.
\newblock \bibinfo{publisher}{Microsoft Press}, \bibinfo{address}{Redmond, WA}.
\newblock
\showISBNx{978-0-7356-1967-8}
\urldef\tempurl%
\url{https://www.safaribooksonline.com/library/view/code-complete-second/0735619670/}
\showURL{%
\tempurl}


\bibitem[Paul~Duvall(2007)]%
        {InfoWorld2007}
\bibfield{author}{\bibinfo{person}{Andrew~Glover Paul~Duvall, Steve~Matyas}.}
  \bibinfo{year}{2007}\natexlab{}.
\newblock \bibinfo{booktitle}{\emph{Introducing continuous integration}}.
\newblock
\urldef\tempurl%
\url{https://www.infoworld.com/article/2077731/introducing-continuous-integration.html?page=3}
\showURL{%
Retrieved june 21, 2022 from \tempurl}


\bibitem[Poncin et~al\mbox{.}(2011)]%
        {MiningRepositories2011}
\bibfield{author}{\bibinfo{person}{Wouter Poncin}, \bibinfo{person}{Alexander
  Serebrenik}, {and} \bibinfo{person}{Mark van~den Brand}.}
  \bibinfo{year}{2011}\natexlab{}.
\newblock \showarticletitle{Process Mining Software Repositories}. In
  \bibinfo{booktitle}{\emph{2011 15th European Conference on Software
  Maintenance and Reengineering}}. \bibinfo{pages}{5--14}.
\newblock
\urldef\tempurl%
\url{https://doi.org/10.1109/CSMR.2011.5}
\showDOI{\tempurl}


\bibitem[Sizilio~Nery et~al\mbox{.}(2019)]%
        {GustavoSizilio2019}
\bibfield{author}{\bibinfo{person}{Gustavo Sizilio~Nery},
  \bibinfo{person}{Daniel Alencar~da Costa}, {and} \bibinfo{person}{Uirá
  Kulesza}.} \bibinfo{year}{2019}\natexlab{}.
\newblock \showarticletitle{An Empirical Study of the Relationship between
  Continuous Integration and Test Code Evolution}. In
  \bibinfo{booktitle}{\emph{2019 IEEE International Conference on Software
  Maintenance and Evolution (ICSME)}}. \bibinfo{pages}{426--436}.
\newblock
\urldef\tempurl%
\url{https://doi.org/10.1109/ICSME.2019.00075}
\showDOI{\tempurl}


\bibitem[Soares et~al\mbox{.}(2021)]%
        {EliezioRevisaoSistematica2022}
\bibfield{author}{\bibinfo{person}{Eliezio Soares}, \bibinfo{person}{Gustavo
  Siz{\'{\i}}lio}, \bibinfo{person}{Jadson Santos},
  \bibinfo{person}{Daniel~Alencar da Costa}, {and} \bibinfo{person}{Uir{\'{a}}
  Kulesza}.} \bibinfo{year}{2021}\natexlab{}.
\newblock \showarticletitle{The Effects of Continuous Integration on Software
  Development: a Systematic Literature Review}.
\newblock \bibinfo{journal}{\emph{CoRR}}  \bibinfo{volume}{abs/2103.05451}
  (\bibinfo{year}{2021}).
\newblock
\showeprint[arXiv]{2103.05451}
\urldef\tempurl%
\url{https://arxiv.org/abs/2103.05451}
\showURL{%
\tempurl}


\bibitem[Statistics(2021)]%
        {SpearmanTest2021}
\bibfield{author}{\bibinfo{person}{Laerd Statistics}.}
  \bibinfo{year}{2021}\natexlab{}.
\newblock \bibinfo{booktitle}{\emph{Spearman's Rank-Order Correlation}}.
\newblock
\urldef\tempurl%
\url{https://statistics.laerd.com/statistical-guides/spearmans-rank-order-correlation-statistical-guide.php}
\showURL{%
Retrieved december 02, 2021 from \tempurl}


\bibitem[Ståhl and Bosch(2014)]%
        {STAHL201448}
\bibfield{author}{\bibinfo{person}{Daniel Ståhl} {and} \bibinfo{person}{Jan
  Bosch}.} \bibinfo{year}{2014}\natexlab{}.
\newblock \showarticletitle{Modeling continuous integration practice
  differences in industry software development}.
\newblock \bibinfo{journal}{\emph{Journal of Systems and Software}}
  \bibinfo{volume}{87} (\bibinfo{year}{2014}).
\newblock
\showISSN{0164-1212}
\urldef\tempurl%
\url{https://doi.org/10.1016/j.jss.2013.08.032}
\showDOI{\tempurl}


\bibitem[tools for high-throughput~data analysis(2021a)]%
        {CorrelationTest2021}
\bibfield{author}{\bibinfo{person}{STHDA~Statistical tools for
  high-throughput~data analysis}.} \bibinfo{year}{2021}\natexlab{a}.
\newblock \bibinfo{booktitle}{\emph{Correlation Test Between Two Variables in
  R}}.
\newblock
\urldef\tempurl%
\url{http://www.sthda.com/english/wiki/wiki.php?id_contents=7307}
\showURL{%
Retrieved december 02, 2021 from \tempurl}


\bibitem[tools for high-throughput~data analysis(2021b)]%
        {ShapiroTest2021}
\bibfield{author}{\bibinfo{person}{STHDA~Statistical tools for
  high-throughput~data analysis}.} \bibinfo{year}{2021}\natexlab{b}.
\newblock \bibinfo{booktitle}{\emph{Normality Test in R}}.
\newblock
\urldef\tempurl%
\url{http://www.sthda.com/english/wiki/normality-test-in-r}
\showURL{%
Retrieved december 02, 2021 from \tempurl}


\bibitem[Vasilescu et~al\mbox{.}(2015)]%
        {VasilescuQualityProdOutcomes2015}
\bibfield{author}{\bibinfo{person}{Bogdan Vasilescu}, \bibinfo{person}{Yue Yu},
  \bibinfo{person}{Huaimin Wang}, \bibinfo{person}{Premkumar Devanbu}, {and}
  \bibinfo{person}{Vladimir Filkov}.} \bibinfo{year}{2015}\natexlab{}.
\newblock \showarticletitle{Quality and Productivity Outcomes Relating to
  Continuous Integration in GitHub}. In \bibinfo{booktitle}{\emph{Proceedings
  of the 2015 10th Joint Meeting on Foundations of Software Engineering}}.
  \bibinfo{publisher}{Association for Computing Machinery},
  \bibinfo{address}{New York, NY, USA}, \bibinfo{pages}{805–816}.
\newblock
\showISBNx{9781450336758}
\urldef\tempurl%
\url{https://doi.org/10.1145/2786805.2786850}
\showDOI{\tempurl}


\bibitem[Vassallo et~al\mbox{.}(2018)]%
        {Vassallo01}
\bibfield{author}{\bibinfo{person}{Carmine Vassallo}, \bibinfo{person}{Fabio
  Palomba}, \bibinfo{person}{Alberto Bacchelli}, {and}
  \bibinfo{person}{Harald~C. Gall}.} \bibinfo{year}{2018}\natexlab{}.
\newblock \showarticletitle{Continuous Code Quality: Are We (Really) Doing
  That?}. In \bibinfo{booktitle}{\emph{Proceedings of the 33rd ACM/IEEE
  International Conference on Automated Software Engineering}} (Montpellier,
  France) \emph{(\bibinfo{series}{ASE 2018})}. \bibinfo{publisher}{Association
  for Computing Machinery}, \bibinfo{address}{New York, NY, USA},
  \bibinfo{pages}{790–795}.
\newblock
\showISBNx{9781450359375}
\urldef\tempurl%
\url{https://doi.org/10.1145/3238147.3240729}
\showDOI{\tempurl}


\bibitem[Widder et~al\mbox{.}(2019)]%
        {ReplicationCIPainPoints2019}
\bibfield{author}{\bibinfo{person}{David~Gray Widder}, \bibinfo{person}{Michael
  Hilton}, \bibinfo{person}{Christian K\"{a}stner}, {and}
  \bibinfo{person}{Bogdan Vasilescu}.} \bibinfo{year}{2019}\natexlab{}.
\newblock \showarticletitle{A Conceptual Replication of Continuous Integration
  Pain Points in the Context of Travis CI}. \bibinfo{publisher}{Association for
  Computing Machinery}, \bibinfo{address}{New York, NY, USA}.
\newblock
\showISBNx{9781450355728}
\urldef\tempurl%
\url{https://doi.org/10.1145/3338906.3338922}
\showDOI{\tempurl}


\bibitem[Yamashita et~al\mbox{.}(2014)]%
        {yamashita2014magnet}
\bibfield{author}{\bibinfo{person}{Kazuhiro Yamashita}, \bibinfo{person}{Shane
  McIntosh}, \bibinfo{person}{Yasutaka Kamei}, {and} \bibinfo{person}{Naoyasu
  Ubayashi}.} \bibinfo{year}{2014}\natexlab{}.
\newblock \showarticletitle{Magnet or sticky? an oss project-by-project
  typology}. In \bibinfo{booktitle}{\emph{Proceedings of the 11th working
  conference on mining software repositories}}. \bibinfo{pages}{344--347}.
\newblock


\bibitem[Zhao et~al\mbox{.}(2017)]%
        {VasilescuImpactCI2017}
\bibfield{author}{\bibinfo{person}{Yangyang Zhao}, \bibinfo{person}{Alexander
  Serebrenik}, \bibinfo{person}{Yuming Zhou}, \bibinfo{person}{Vladimir
  Filkov}, {and} \bibinfo{person}{Bogdan Vasilescu}.}
  \bibinfo{year}{2017}\natexlab{}.
\newblock \showarticletitle{The impact of continuous integration on other
  software development practices: A large-scale empirical study}. In
  \bibinfo{booktitle}{\emph{2017 32nd IEEE/ACM International Conference on
  Automated Software Engineering (ASE)}}. \bibinfo{pages}{60--71}.
\newblock
\urldef\tempurl%
\url{https://doi.org/10.1109/ASE.2017.8115619}
\showDOI{\tempurl}


\end{thebibliography}


\end{document}